\documentclass[graybox]{svmult}

\usepackage{mathptmx}       
\usepackage{helvet}         
\usepackage{courier}        
\usepackage{type1cm}        
\usepackage{makeidx}         
\usepackage{graphicx}        
\usepackage{multicol}        
\usepackage[bottom]{footmisc}

\makeindex             

\begin{document}

\title*{The balance of power: accretion and feedback in stellar mass black holes}
\author{Rob Fender and Teo Mu\~noz-Darias}
\institute{Rob Fender \at Astrophysics, Department of Physics, University of Oxford,
e-mail: rob.fender@astro.ox.ac.uk \and Teo Mu\~noz-Darias \at Astrophysics, Department of Physics, University of Oxford; Instituto de Astrof\'isica de Canarias; Universidad de La Laguna, Departamento de Astrof\'isica, 
e-mail: teo.munoz-darias@iac.es}

%
%
\maketitle

\abstract{In this review we discuss the population of stellar-mass black holes in our galaxy and beyond, which are the extreme endpoints of massive star evolution. In particular we focus on how we can attempt to balance the available accretion energy with feedback to the environment via radiation, jets and winds, considering also possible contributions to the energy balance from black hole spin and advection. We review quantitatively the methods which are used to estimate these quantities, regardless of the details of the astrophysics close to the black hole. Once these methods have been outlined, we work through an outburst of a black hole X-ray binary system, estimating the flow of mass and energy through the different accretion rates and states.
While we focus on feedback from stellar mass black holes in X-ray binary systems, we also consider the applicability of what we have learned to supermassive black holes in active galactic nuclei. As an important control sample we also review the coupling between accretion and feedback in neutron stars, and show that it is very similar to that observed in black holes, which strongly constrains how much of the astrophysics of feedback can be unique to black holes.}

\section{Introduction}
\label{intro}

Stellar mass black holes are the final phases of the evolution of the most massive stars, which runs quickly (in a million years or less) from protostellar gas cloud, through a brief main sequence to a final collapse (which may or may not be accompanied by a supernova explosion). Many millions of such stellar remnants are expected to exist in our galaxy alone, the vast majority of which remain undiscovered. Only a tiny subset, the tip of the tip of the iceberg, are revealed to us via the extreme luminosities which can be produced via accretion. 

These stellar mass black holes are the sites of the most extreme gravitational curvature in the present-day universe, and potentially the best tests of general relativity available to us (Psaltis 2008).  However, in this review we will concentrate on what are the outcomes of the accretion process, rather than the details of the astrophysics which occur close to the event horizon. Somehow, the black hole contrives to feed back to the surrounding universe a large fraction of the energy it could potentially have swallowed, in this way acting to heat its environment rather than acting as a sink. We can attempt to balance the budget of spin, fuel, advection, kinetic and radiative feedback within a black box which contains the black hole and its accretion flow, to try and discern the patterns which occur there. This is important across a wide range of scales, comparable to the mass range of black holes themselves: from heating of our local environment (there must be isolated black holes and neutron stars within a few parsecs of us doing just this), to the feedback from AGN which, it seems, regulated the growth of the most massive galaxies.

The outline of this review is as follows. We begin with a very simple introduction to accretion, followed by an up to date picture of the phenomenology of accretion and feedback associated with stellar mass black holes accreting in binary systems. We will take the reader through the various methods used to try and estimate the power associated with different forms of feedback (steady and transient jets, accretion disc winds), and then run through the course of an X-ray binary outburst to see how feedback varies and how the accumulated flavours of feedback stack up against each other. We shall discuss the evidence that what we have learned from stellar mass black holes may be applicable to AGN, which may be important for galaxy formation. Finally, we shall wrap up with an up to date summary of the relation between accretion and feedback in neutron stars, and show that it very closely resembles that observed in black holes.

\subsection{Accretion}

Accretion is the process whereby an object captures matter as a result of its gravitational attraction. The accreted matter has therefore necessarily fallen into a deeper gravitational potential and energy is liberated. In the simplest case of an object initially at rest and left to fall towards the gravitating object, the potential energy converts to kinetic energy. If the central accreting object has, like nearly everything in the universe, a surface, then this kinetic energy must convert into other forms when the object is stopped at the surface. In the case of a black hole, our simple infalling object would just cross the event horizon and vanish from our part of the universe, advecting this kinetic energy.

\begin{figure}
\includegraphics[scale=.42]{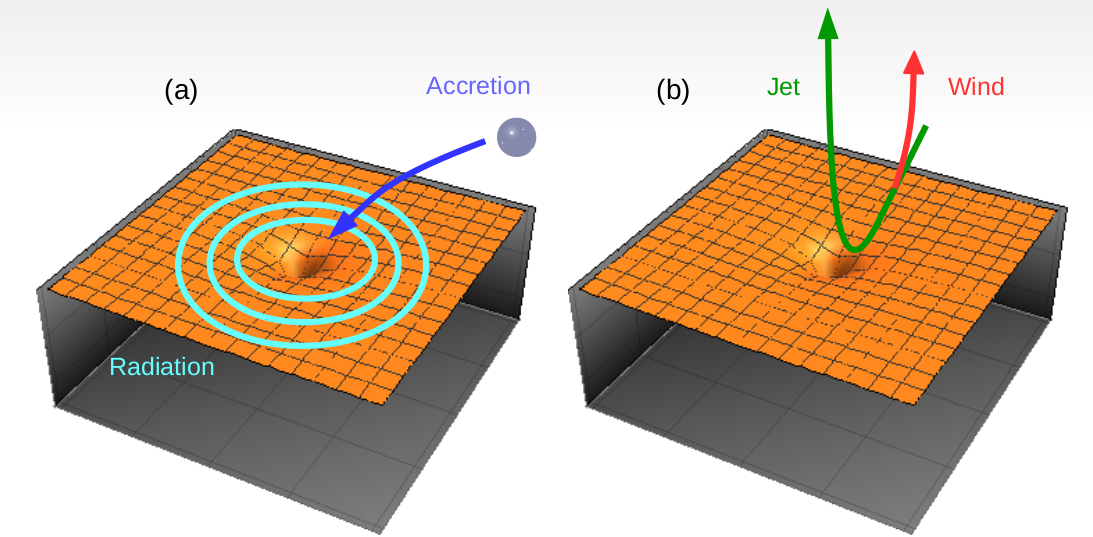}
\caption{Accretion and ejection. The gridded surface illustrates a simple Newtonian gravitational potential. In (a) mass is accreted onto a central object, which results in the (isotropic, usually) release of radiation to infinity. However, this liberated accretion energy can also feed back to the environment via jets and winds. In (b) we illustrate (more stylistically than realistically) the (anisotropic) loss of mass and energy in a jet from close to the central accretor, or in a wind further out. If ejected mass is not to fall back, it needs to exceed the local escape velocity, which sets a minimum speed. As a result, in order to eject all of the incoming mass back out to infinity you would need a perfect engine and no other release of energy e.g. radiation. Much more likely is either/both of a jet with a relatively low (compared to the inwards accretion rate) mass loading, or ejection of significant mass from further out in the gravitational potential (wind), powered by the central accretion of mass.
}
\label{potential}  
\end{figure}

The energy change per unit mass in falling from infinity to a radius R outside of a gravitating mass M is:

\[
\Delta E = -\frac{GM}{R} + \frac{GM}{\infty} = -\frac{GM}{R}
\]

This demonstrates that the amount of liberated energy is proportional to the ratio of mass to radius: more compact $\rightarrow$ more efficient accretion. Fig \ref{potential} illustrates, schematically and extremely simply, the release of energy via accretion, and the resulting feedback to the environment in terms of radiation, jets and winds.

However, nearly all matter in the universe has angular momentum, which means that it will not fall directly into the black hole. Simple thought experiments, detailed theory, and numerical simulations suggest that in fact an accretion disc will form, in which the accreting matter moves in to the central accretor via a set of essentially circular orbits. Direct observational evidence for such discs exists in the form of spatially and spectrally resolved rotating flows around supermassive and stellar mass black holes. For a sample of the extensive background on this subject the reader is directed to e.g. Pringle \& Rees (1972), Shakura \& Sunyaev (1973), Papaloizou \& Lin (1995), Lin \& Papaloizou (1996), Balbus (2003), references therein and citations thereof.

\subsection{Stellar mass black holes}

\begin{figure}
\includegraphics[scale=.65]{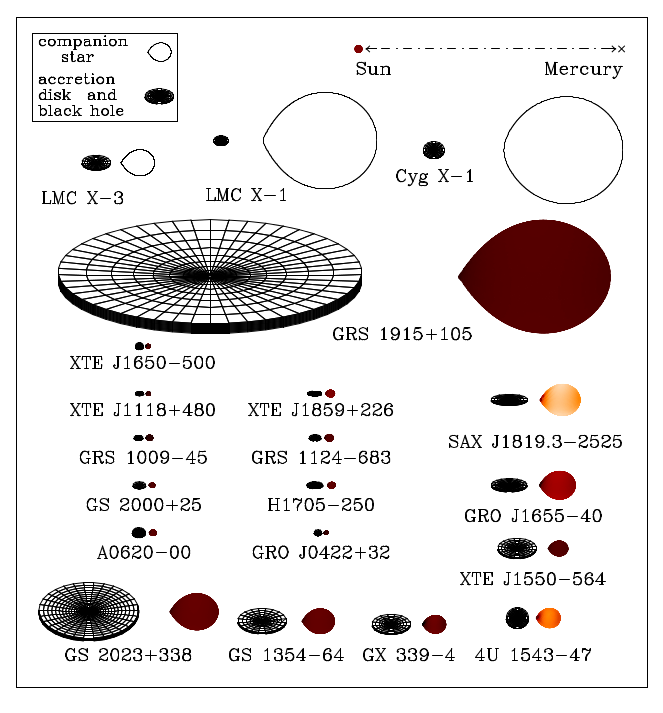}
\caption{Twenty black hole binary systems scaled to our best current estimates of their physical dimensions. Credit: Jerry Orosz (see http://mintaka.sdsu.edu/faculty/orosz/web/).}
\label{orosz}  
\end{figure}

The term stellar mass black holes refers to black holes in the mass range from a few to perhaps 20 solar masses, i.e. those which could be direct endpoints of the evolution of a single massive star and whose masses do not represent any significant accumulation of mass via accretion or mergers (unlike, probably, many supermassive black holes in active galactic nuclei). 

The best evidence for stellar mass black holes comes from optical spectroscopy of low mass X-ray binaries (where the {\em low mass} refers to the mass of the non-compact mass donor, being typically less than one solar mass) in quiescence, when the accretion disc is faint enough to allow the donor star to be detected and its radial velocity variations at the orbital period revealed. In several cases the mass function, an absolute minimum mass for the accretor, is well in excess of $3M_{\odot}$, providing strong evidence for the existence of black holes. For a subset of these systems, a detailed modelling of their X-ray and/or optical fluxes as a function of time (light-curves) allows estimation of the inclination of the binary, and subsequent constraint of the black hole mass with an uncertainty as small as 10\% (e.g. Casares \& Jonker 2014). This method has yielded nearly 20 objects known as \textit{dynamical black holes}; estimated physical dimensions of these are presented in Fig {\ref{orosz}}. Another $\sim 40$ {\em black hole candidates}, share a strong phenomenology with the previous class, which is different from that observed from systems containing neutron stars, typically in the X-ray spectral and variability properties (e.g. van der Klis 2005, Mu\~noz-Darias et al. 2014). We shall refer to these systems collectively as the black hole X-ray binaries (BHXRBs).

The black hole discovery rate is currently limited by the (X-ray) outburst rate of black hole binaries, which is very small ($\sim 2$ per year) due to their low duty cycles (fraction of time they are X-ray active, typically perhaps 1\%). Therefore, the above number ($\sim 60$) only accounts for a tiny fraction of the actual black hole binary galactic population, thought to be above $\sim 10^4$ objects (e.g. Yungelson et al. 2006). They, similarly, only represent a small amount out of the total number of stellar mass black holes harboured in the Milky Way, that has been estimated to be in the range $10^8$--$10^9$ (Van den Heuvel 1992; Brown \& Bethe 1994). The actual number depends on the supernova rate across the life time of our Galaxy and on details of the stellar evolution. Here, an important question to address regards the maximum mass for a stellar remnant to still be a neutron star, and the minimum to form a black hole. There is currently some discussion as to whether or not there may be a mass gap between neutron stars - the most massive known has $\sim 2$M$_{\odot}$ (Demorest et al. 2010) - and stellar mass black holes ($> 4$--5 M$_{\odot}$ so far), but any conclusions drawn are necessarily tentative due to uncertainty about selection effects and the aforementioned, poor statistics (see e.g. discussion in \"{O}zel et al. 2010).

\section{Patterns of disc-jet coupling}
\label{patterns}

\begin{figure}
\includegraphics[scale=.55]{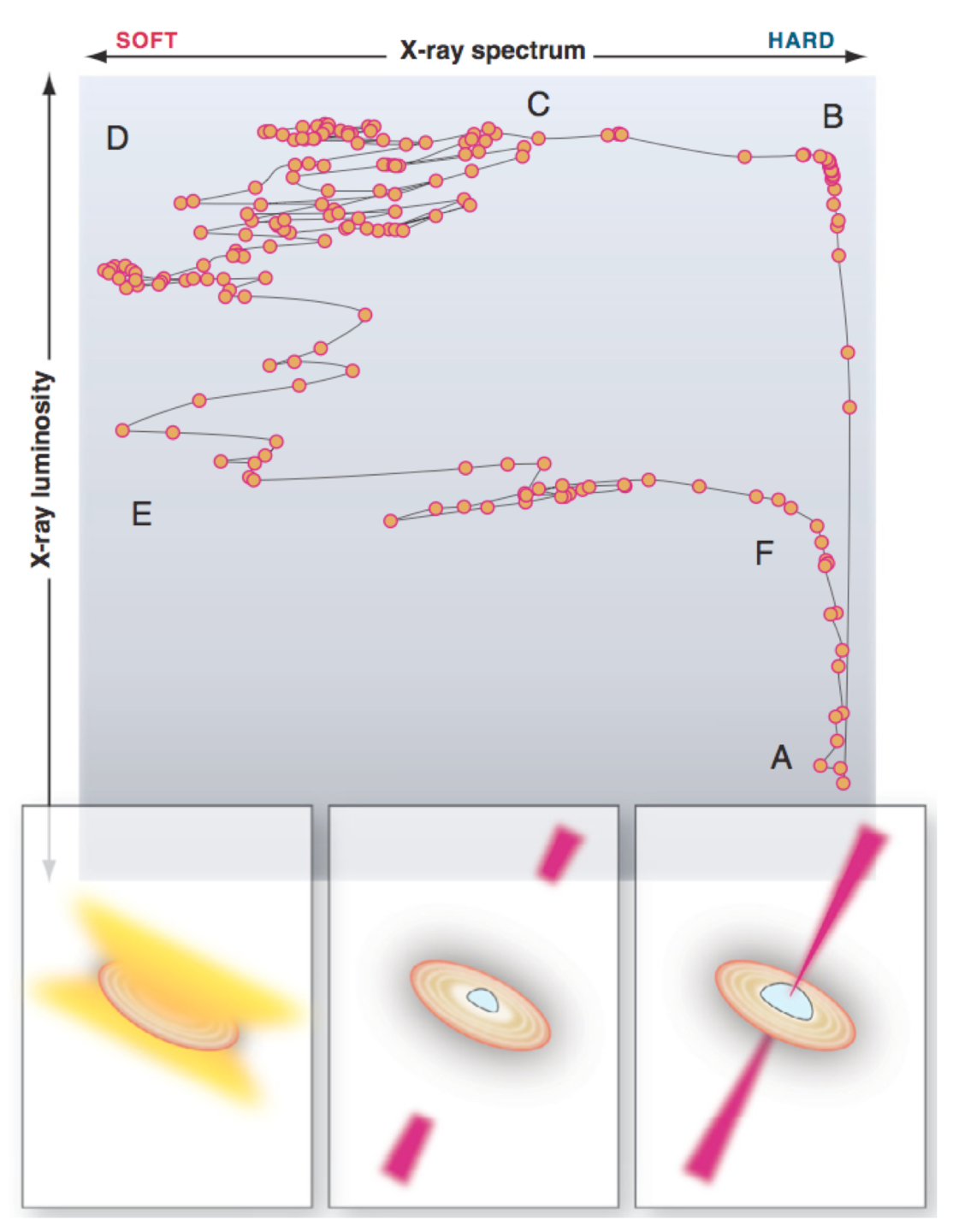}
\caption{Patterns of accretion:outflow in black hole X-ray binaries. The diagram presents X-ray monitoring of the black hole binary GX 339-4, which completed the path A $\rightarrow$ F on a timescale of about one year.
Hard X-ray states (to the right) are associated with (quasi-)steady jet production and little or no winds. Conversely, the soft states to the left are associated with strong accretion disc winds and no strong core jet. Transitions between hard and soft X-ray states are associated with large, sometimes multiple, discrete ejection events. From Fender \& Belloni (2012), to which the interested reader is directed for more references in this area, and which combined primarily the phenomenology first described in Fender, Belloni \& Gallo (2004) and Ponti et al. (2012).}
\label{FB2012}
\end{figure}

The clear phenomenological connection between accretion and outflow in BHXRBS has been detailed in many reviews over the past few years, and we refer the interested reader to Fender \& Belloni (2012) and references therein. In short, there are a small number of accretion `states' which correspond to patterns of behaviour in X-ray spectra and timing properties (as revealed in Fourier power spectra), and these seem to have a well defined connection to modes of outflow; see Figs \ref{FB2012}, \ref{zhang}. 

As a function of time, an outburst progresses through the following stages:

\begin{itemize}
\item{A `hard' state, in which the X-ray spectrum is dominated by a component which peaks around 100 keV is likely to originate in thermal comptonisation, although there may be some contribution from jet (or other) synchrotron emission.
This state, associated with a powerful, quasi-steady jet, is associated with the initial rise of the transient}
\item{At some X-ray luminosity, typically $\geq$ 10\% of the Eddington luminosity, the source passes briefly through a phase or phases of transient discrete ejections, and progresses to the `soft' state. This state has an X-ray spectrum which peaks at about 1 keV, probably arising from an optically thick accretion disc.}
\item{In the `soft' state there is a much weaker or non-existent core jet and a strong accretion disc wind.} 
\item{Eventually (on average a handful of months later) the outburst fades and the system switches back to the hard state, typically around 1\% of the Eddington X-ray luminosity, and the wind switches off and the jet back on.}
\end{itemize}

The jet phenomenology described above is rather well established via a large (100s) number of radio observations, although many details remain to be filled in. However, much more needs to be established about the duty cycle of the accretion disc wind in the soft state, which is crucial to the calculations outlined next.

\section{Feedback}

The balance between black hole spin, the available accretion energy, advection, and the radiative and kinetic feedback can be summarised as:

\begin{equation}
P_{spin} + P_{accretion} = L_{bolometric} + L_{kinetic} - L_{advected}
\label{balance}
\end{equation}

Where $P_{spin}$ and $P_{accretion}$ are the `input' powers provided at any given moment (which may change throughout an outburst) from the black hole spin and accretion of mass, respectively. On the other side of the equation we have the `sink' terms for this power: $L_{bolometric}$ is the bolometric radiative output, $L_{kinetic}$ is the feedback of kinetic energy to the environment (via jets and winds), $L_{advected}$ measures the amount of available power which is actually advected across the black hole event horizon. Obviously, $L_{advected}$ is a term unique to black holes. We further note that while $P_{accretion} \geq 0$, the term $P_{spin}$ can be positive or negative. While commonly considered in terms of a rapidly spinning black hole having `extra' power available to put into e.g. a jet, a non-rotating black hole which begins to accrete will begin to {\em gain} spin energy, and act as a sink for the available power. Spin-up and spin-down of neutron stars is routinely observed in observations of X-ray pulsars (e.g. Bildsten et al. 1997).
The equation, being essentially a statement of the conservation of energy, is reminiscent of the first law of thermodynamics (and is equally susceptible to the addition of further terms).

As is common practice, we can write the available accretion power as 

\begin{equation}
P_{accretion} = \frac{G M \dot{m}}{r} = \eta \dot{m} c^2
\end{equation}

Where $\eta$ parametrises the accretion efficiency and $\dot{m}$ is the accretion rate.
The total kinetic feedback can be crudely divided into that associated with (relativistic) jets and (non-relativistic) winds, since these appear to be the main modes, apparently mutually exclusive, in black hole X-ray binaries:

\begin{equation}
L_{kinetic} = L_{jet} + L_{wind}
\end{equation}

Thus an expanded form of equation \ref{balance} can be written as:

\begin{equation}
P_{spin} + \eta \dot{m} c^2 = L_{bolometric} + L_{jet} + L_{wind} - L_{advected}
\label{balance2}
\end{equation}

While the details of the inner accretion flow, including the jet formation region, can potentially probe some of the most exciting astrophysics in the entire universe, they are extremely hard to measure via the approaches outlined in this review. However, what we can do is try to balance equations (\ref{balance}, \ref{balance2}) to see how black holes take available accretion energy and convert it into radiation and kinetic power which affects the environment. As we shall see, we have the tools to estimate a number of the above terms, and can make good progress towards understanding the balance of power in accreting black hole. However, the fact that some of the terms are not independent (in the ways they are estimated), and others evade easy estimation, means that we are a long way from a comprehensive solution to this energy balance.

\subsection{Radiation}

\begin{figure}
\begin{center}
\includegraphics[scale=0.45]{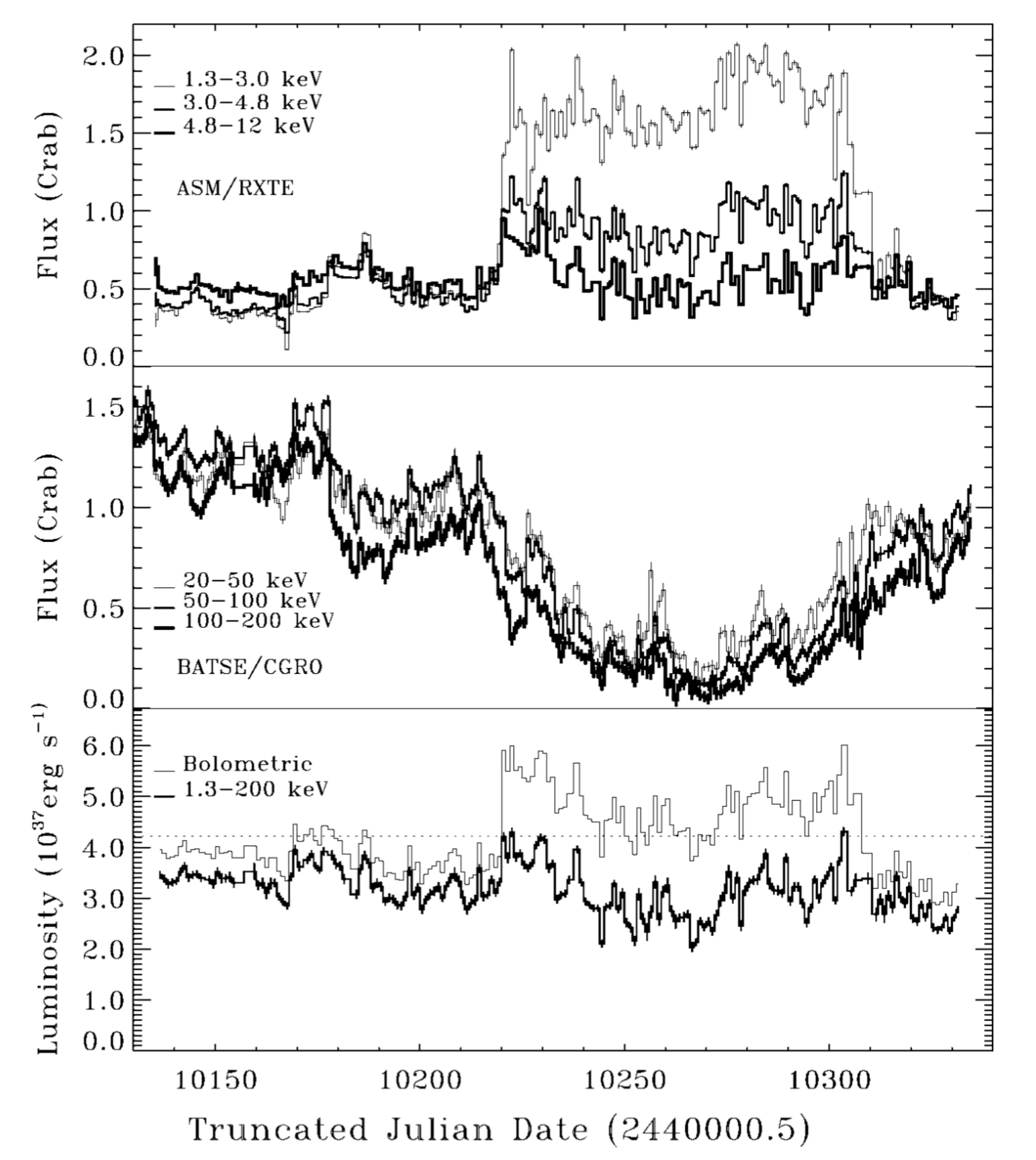}
\caption{Soft- and hard-X-ray, as well as (estimated) bolometric flux from Cygnus X-1 across state transitions in 1996. Importantly, although the source is at $\sim 2$\% Eddington luminosity, there is still less than a 50\% change in luminosity across the transition, implying that the hard state is radiatively efficient. From Zhang et al. (1997).}
\label{zhang}
\end{center}
\end{figure}

This is in principle the easiest of the quantities to measure, at least at high accretion rates (more than say 1\% of the Eddington accretion rate), where the radiative output appears to be dominated by X-ray emission (naturally, considering the energy release and size scale -- see Frank, King \& Raine 2002). Bolometric corrections to the 1-10 keV emission have been estimated to be $\sim 2$ in the `soft' X-ray state and $\sim 5$ in the `hard' state (e.g. Migliari \& Fender 2006). Considerable uncertainties exist regarding exactly how this bolometric correction may change throughout an outburst, as the high-energy cutoff (typically $\sim 100$ keV in the hard state) may evolve significantly. On the other hand, the well-measured bolometric luminosity of Cyg X-1 appears to change by less than 50\% in the 1.3-200 keV band during a hard $\rightarrow$ soft(/intermediate) state transition (Zhang et al. 1997; see Fig \ref{zhang}). 

\subsubsection{On radiative efficiency}

While it is generally accepted that the soft, disc-dominated, accretion states should be radiatively efficient, with $\eta \sim 0.1$, it is far less clear how radiatively efficient the hard states are. A common interpretation of the hard (hot) states is that they are advection-dominated, in which much of the available accretion energy is `trapped' in hot ions which cross the event horizon before they can radiate (e.g. Ichimaru 1977; Esin, McClintock \& Narayn 1977; Feng \& Narayan 2014 and references therein). Interestingly (see below), such flows have been for a long time tentatively associated with relativistic jets (e.g. Rees et al. 1982). In such flows the radiative efficiency may drop with the accretion rate, such that 

\begin{equation}
L_X \propto \dot{m}^c
\end{equation}

where $c > 1$. A commonly accepted value is $c \sim 2$, which fits nicely into pictures of accretion, radiative efficiency and feedback, considering the observed radio:X-ray correlations in black hole (potentially inefficient) and neutron star (necessarily efficient) binaries (see below as well as discussions in e.g. Fender, Gallo \& Jonker 2003; Coriat et al. 2011). However, the lack of a dramatic jump in luminosity at the transition from the hard (hot) to soft (cool) accretion flows, at least in Cyg X-1 (see previous section; Fig \ref{zhang}) already tells us that the radiative efficiency of the states must be comparable (at least in this case, and at this luminosity). This can lead to difficulties in connecting the accretion rate smoothly through an outburst (e.g. Russell et al. 2007 and later in this review).

One additional frustration for observers is that at accretion rates below $\sim 1$\% (of the Eddington luminosity, in X-rays), the peak of the emission from the accretion disc moves into the UV part of the spectrum, which is extremely hard to measure for nearly all galactic X-ray binaries, as it corresponds to a peak in the absorption and scattering of photons by dust in the interstellar medium (but see the case of XTE J1118+480 where observations were made: Hynes et al. 2003). 

In summary, there is a lot of uncertainty as to how the radiative efficiency evolves between the peak of an outburst and quiescence. In evaluating the black hole outburst later in the review, we shall consider both radiatively efficient ($c=1$) and inefficient (specifically, $c=2$) cases.

\subsection{Jets}

Jets from black hole X-ray binaries (BHXRBs), and indeed most classes of accreting object, are mainly revealed via their radio emission. This is not necessarily because they do not emit in other bands which are well observed, e.g. optical, X-ray, but that in those bands it is hard to distinguish jet emission from that produced by other processes (e.g. stars, accretion flows). 

In black hole X-ray binaries the radio emission has essentially three `modes', whose connection to the accretion state is phenomenologically clear (see section \ref{patterns}), but can be summarised as:

\begin{itemize}
\item{{\bf Steady jet:} in the hard state, the radio emission appears quasi-steady, with a flat spectrum}
\item{{\bf Transient ejections:} during hard $\rightarrow$ soft state transitions, radio flares are observed which have been, in several cases, spatially resolved in blobs moving away from the central source with bulk Lorentz factor $\Gamma \geq 2$.}
\item{{\bf Off:} in the soft state, the core radio jet appears to be completely off, with non-detections a factor of $\sim 100$ below the expected level for the hard state (the best example is in Russell et al. 2011)}
\end{itemize}

In the following subsections we shall outline how to estimate the kinetic power associated with each of these types of radio emission.

\subsubsection{Steady jets}

\begin{figure}
\includegraphics[scale=.65]{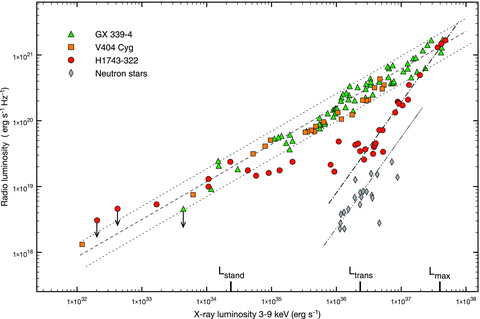}
\caption{The correlation between GHz radio flux density and 3--9 keV X-ray flux in the black hole binaries GX 339-4, V404 Cyg and H1743-322 (from Coriat et al. 2012). GX 339-4 and V404 show a very clear correlation which for GX339-4 has been observed to be repeat very consistently between multiple different hard states, indicating a remarkable stability in jet formation (see also Corbel et al. 2013 for the complete data set on this source). H1743-322 is a clear example of a `radio quiet' population which appear to occupy a lower, steeper, track in the plane at high luminosities, possibly rejoining the `radio loud' track at $L_X \sim 0.1$\% Eddington. This track is close to the region of the diagram for neutron stars, both in normalisation and slope (caveat rather few points for the neutron stars). The underlying reason for two separate branches remains a mystery.}
\label{coriat}
\end{figure}

In the hard X-ray state, flat-spectrum radio emission is seen whose flux density at GHz frequencies scales with the 1-10 keV X-ray luminosity roughly as

\begin{equation}
L_{radio} \propto L_X^b
\end{equation}

where $0.6 \leq b \leq 0.7$ for the best-sampled sources, in particular for the very well sampled source GX 339-4, which is the subject of our detailed study later. Corbel et al. (2013) reports the state of the art for GX 339-4, which follows a correlation with $b=0.62 \pm 0.01$. Figure \ref{coriat} shows the radio:X-ray correlation for three black hole binaries, including GX 339-4, as well as a much smaller sample of neutron stars. Very obvious is the fact that, while GX 339-4 and V404 Cyg show a nice power-law relation, in the black hole binary H1743-322 the situation is more complex. The reason behind the `radio quiet' branch which H1743-322 appears to follow at X-ray luminosities above about $10^{36}$ erg s$^{-1}$ is at present completely unclear, but this phenomena has been seen in an increasing number of sources. 

The key issue is the relation between a measured core (generally unresolved) radio luminosity and the associated kinetic power, something which has been much explored not only for BHXRBs but also for the radio cores of AGN. K\"ording et al. (2006a) presented an approximate relation, calibrated on X-ray binaries but also (tentatively) applied to the cores of some low Eddington ratio AGN:

\begin{equation}
L_{jet} \sim 4 \times 10^{36} \left(\frac{L_{8.6}}{10^{30}}\right)^{12/17}
\label{steadypower}
\end{equation}

where $L_{8.6}$ is the radio luminosity calculated as 

\begin{equation}
L_{8.6} = 4 \pi d^2 \nu_{8.6}  F_{8.6}
\end{equation}

where $d$ is the distance to the source and $\nu_{8.6} = 8.6 \times 10^9$ is the frequency associated with 8.6 GHz. In other words, this is the radio luminosity assuming a flat spectrum to low frequencies.

As discussed in K\"ording et al. (2006a), this function is quite similar to those derived by other means. As an example, in a recent review, Merloni \& Heinz (2012) present an almost identical relation between radio core luminosity and inferred jet kinetic power,

\begin{equation}
L_{jet} \sim 1.6 \times 10^{36} \left(\frac{L_{radio}}{10^{30}}\right)^{0.81}
\end{equation}

which was based up a sample of AGN for which radio cores could be measured at the same time as analysis of cavities and shocks in the surround medium. In other words, current analyses are consistent with {\em a single relation between core radio luminosity and jet power across the mass range from BHXRBs to AGN}. This is a remarkable result, and is illustrated in Fig \ref{powerfns}.

A quick note on the radio:quiet branch of X-ray binaries such H1743-322 (Fig \ref{coriat}): it may be that the above relations between radio luminosity and kinetic power are entirely appropriate here, meaning that the jet power is rather lower at high luminosities, but then remains (mysteriously) constant while the source declines in X-ray luminosity by almost two orders of magnitude. It may equally be that for some reason in the `radio quiet' zone the relation between radio emission and jet power has deviated from the simple relations given above.

In summary, we use simple relations, such as those outlined above, to estimate the kinetic power associated with a given core 
{\footnote{In general `core' is taken as shorthand for unresolved central (at, or close to, the black hole) radio emission, which typically has a flat spectrum. In a few X-ray binaries, and many active galactic nuclei, the total radio luminosity is much larger when extended jets and lobes are included.}
radio luminosity. It is worth remembering that a lot of hard work has gone into estimating/measuring the normalisations and slopes of samples which result in such an apparently straightforward connection. Something realised early on was that the combination of the kinetic power functions with the observed nonlinear X-ray:radio correlation was consistent with (not proof of) a scenario in which the jet power was scaling linearly with the accretion rate while the X-ray luminosity was scaling approximately as the square of the accretion rate (e.g. Fender, Gallo \& Jonker 2003). This relation between accretion rate and X-ray luminosity fits in turn, roughly, with the scalings proposed for radiatively inefficient accretion flows (e.g. Mahadevan 1997). 

\begin{figure}
\includegraphics[scale=.65]{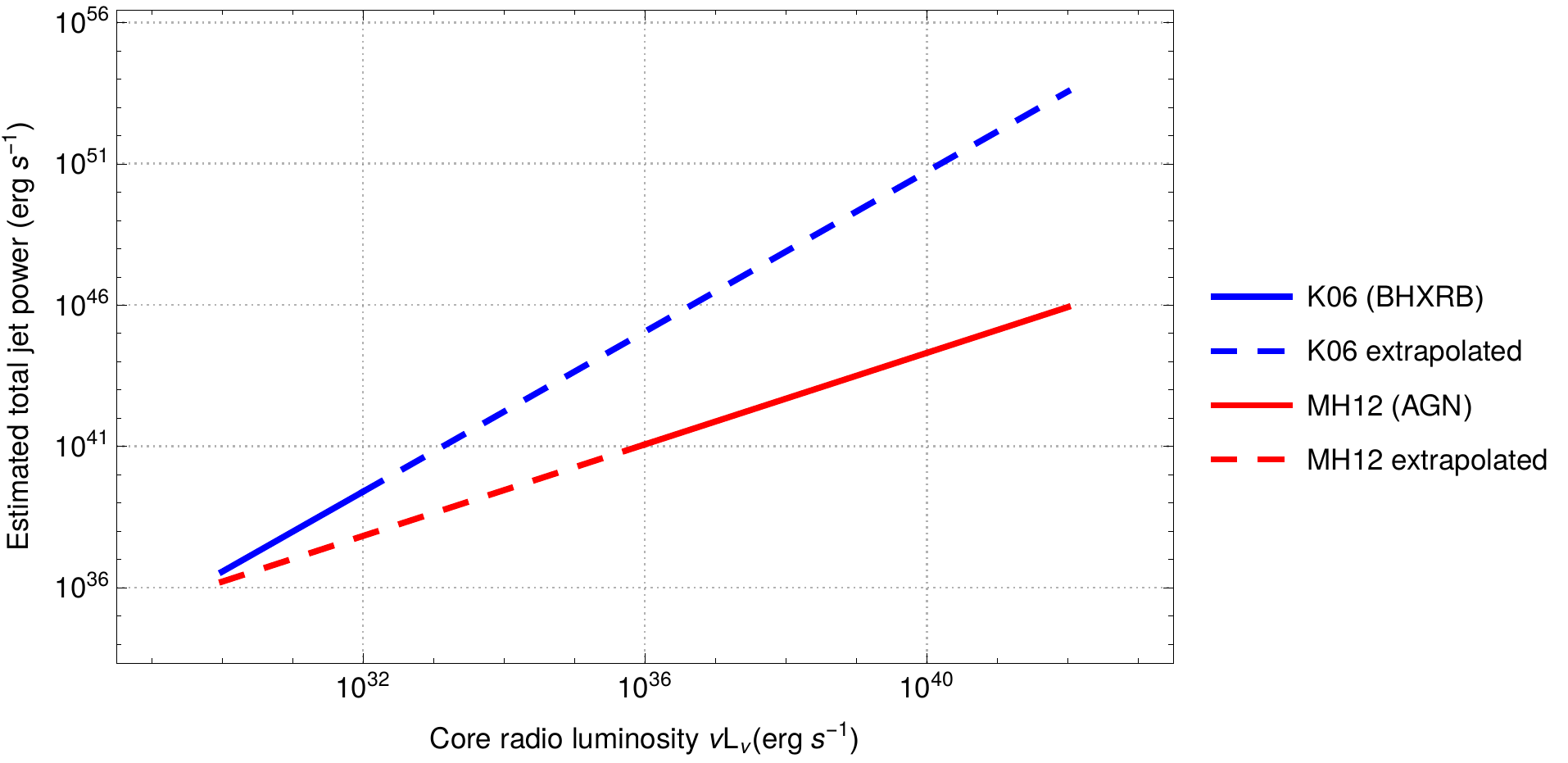}
\caption{Two functions which can be used to estimate the total kinetic power associated with a jet by measuring the core radio luminosity. The function K06 originates in K\"ording et al. (2006a) and was estimated based on observations of X-ray binaries. The function MH12 is from Merloni \& Heinz (2007, 2012) and was estimated  based on observation of AGN with associated radio cavities. For each function the range over which it was evaluated is indicated by the solid line, and the extrapolation by the dashed line. It is remarkable that the functions are broadly similar, across six to seven orders of magnitude in black hole mass.}
\label{powerfns}
\end{figure}

\subsubsection{Transient jets associated with flares}

Major ejection events are associated with bright flares observed in the radio light curve (e.g. Harmon et al. 1995; Fender et al. 1999; Fender, Homan \& Belloni 2009; Miller-Jones et al. 2012). In some cases we can directly image the components as they move away from the central accretor at relativistic bulk velocities (see Fig \ref{1915images} for two examples from the powerful jet source GRS 1915+105).

\begin{figure}
\includegraphics[scale=.5]{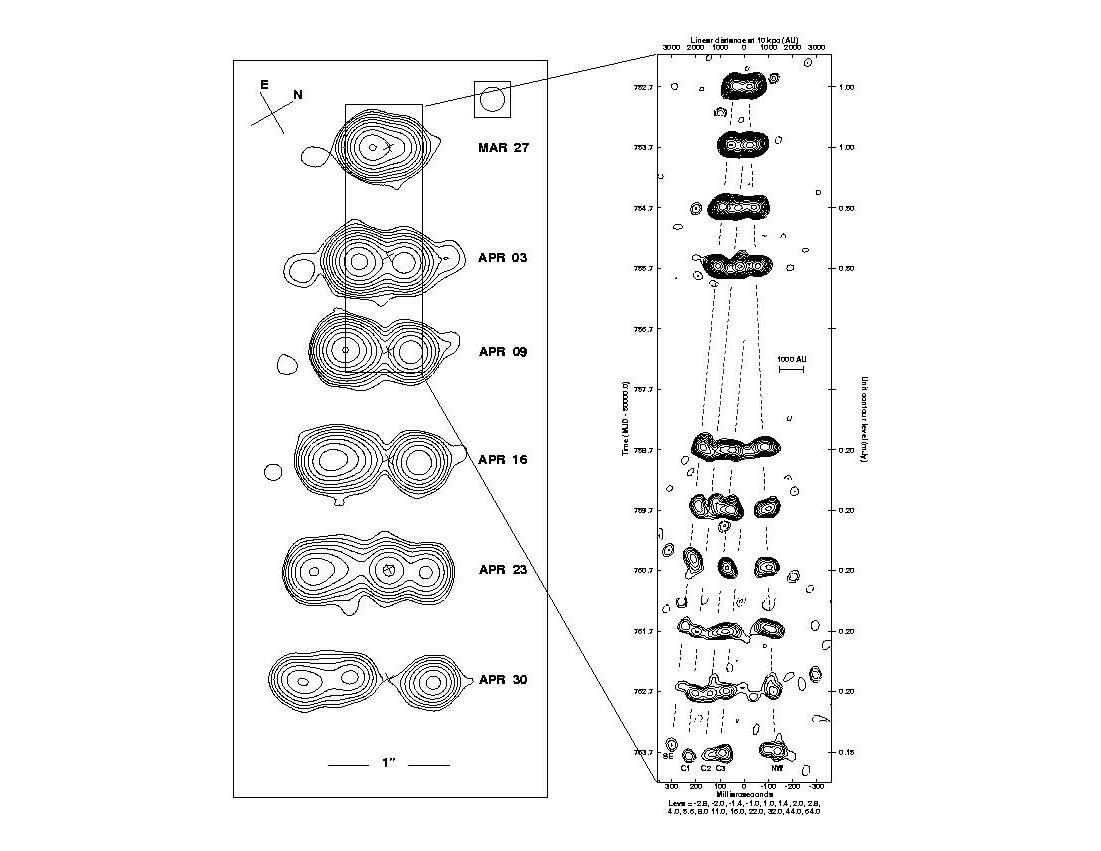}
\caption{Spatially resolved relativistic ejections from the black hole binary GRS 1915+105, a powerful and repeating source of such events since it `switched on' in 1992. This figure is adapted from Mirabel \& Rodriguez (1994: VLA observations, left panel) and Fender et al. (1999: MERLIN observations, right panel). This system shows a remarkably rich phenomenology, and the interested reader is directed to the review of Fender \& Belloni (2004) and subsequent works by e.g. Miller-Jones et al. (2005). }
\label{1915images}
\end{figure}

When we can associate a given synchrotron luminosity with a given volume, we can employ a method for estimating the power during the ejection event which has its roots in calculations made over half a century ago by Burbidge (1959) for the radio lobes of AGN. In essence the total energy (which is the sum of the integrated energies in the electrons and magnetic field) can be minimised as a function of the magnetic field. Since this minimum energy is associated with approximate equipartition of internal energy between electrons and magnetic field, it is refered to as the equipartition magnetic field.

\begin{figure}
\includegraphics[scale=.42]{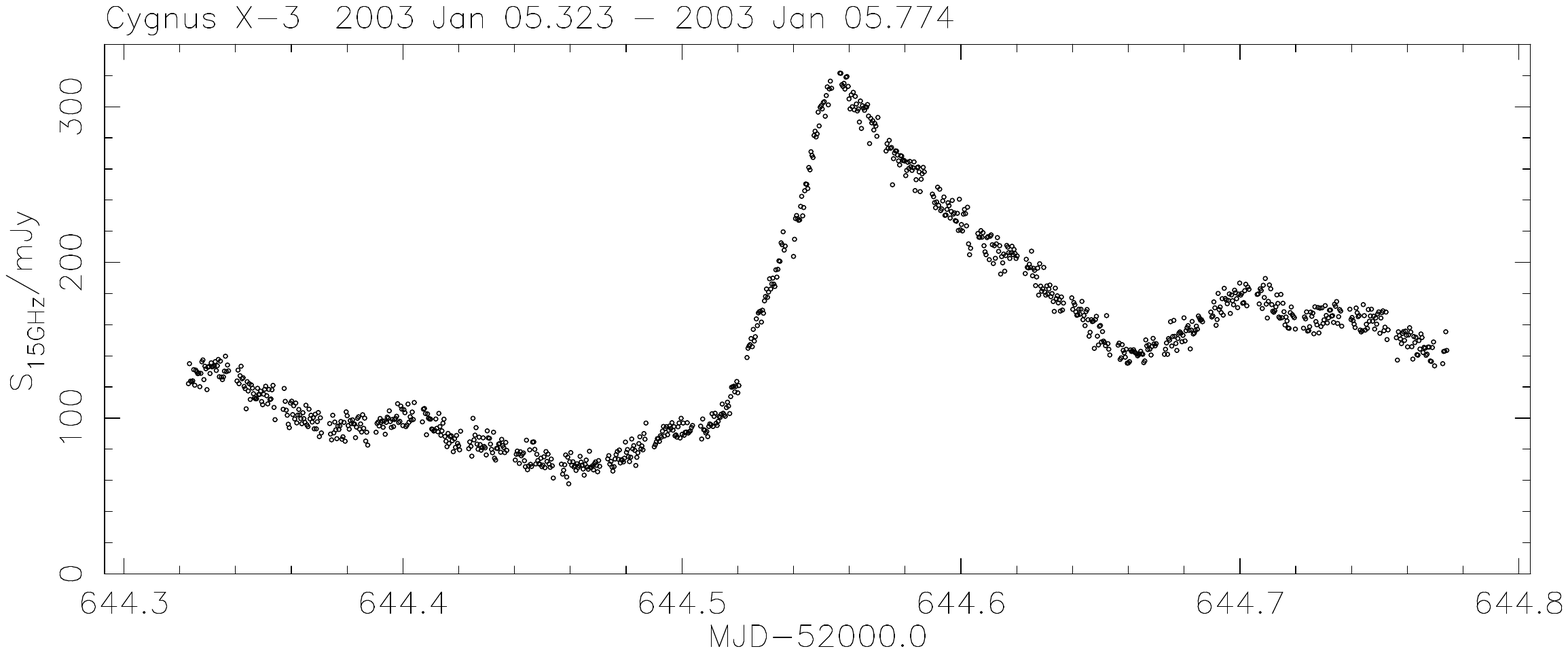}
\caption{Observation of a radio flare event from the jet
source Cyg X-3 at 15 GHz. The rise time of the event $\sim 0.04$ d,
allows an estimation of the size of the region associated with the
event, and thus the minimum energy. Observations from the Ryle
Telescope (Guy Pooley, private communication).}
\label{cygx3flare}
\end{figure}

Longair (1994) gives a clear explanation of the method, and the interested reader is directed there. Repeating some of his useful formulae, a lower limit to the energy associated with a finite volume of synchrotron emitting plasma can be obtained from a simple estimate of the monochromatic luminosity at a given frequency which is associated
with that volume:

\begin{equation}
E_{\rm min} \sim 8\times 10^{6} \eta^{4/7} \left(\frac{V}{{\rm
    cm}^3}\right)^{3/7} \left(\frac{\nu}{\rm Hz}\right)^{2/7}
    \left(\frac{L_{\nu}}{{\rm erg \phantom{0}
    s^{-1} Hz^{-1}}}\right)^{4/7} \phantom{00} {\rm erg}
\end{equation}

where $\eta = (1+\beta)$ and $\beta = \epsilon_{\rm p} / \epsilon_{\rm
 e}$ represents the ratio of energy in protons to that in electrons,
and assuming $p=2$ (where $p$ is the electron distribution index, $N(E)dE \propto E^{-p}$).  
It is generally accepted that $\beta \sim 0$ and
therefore $\eta \sim 1$, but this has not really been observationally proven. 

In the more common case where we do not image the source but rather infer
its size from the rise time $\Delta t$ of an event (i.e. using $V =
(4/3) \pi (c \Delta t)^3$ with a flux density $S_{\nu}$ originating at
an estimated distance $d$, the formula can be rewritten as

\begin{equation}
E_{\rm min} \sim 3\times10^{33} \eta^{4/7} \left(\frac{\Delta t}{\rm
  s}\right)^{9/7} \left(\frac{\nu}{\rm GHz}\right)^{2/7}
\left(\frac{S_{\nu}}{\rm mJy}\right)^{4/7} \left(\frac{d}{\rm
  kpc}\right)^{8/7} \phantom{00} {\rm erg}
\end{equation}

The related mean power into the ejection event:

\begin{equation}
P_{\rm min} = \frac{E_{\rm min}}{\Delta t} \sim 3\times10^{33} \eta^{4/7} \left(\frac{\Delta t}{\rm
  s}\right)^{2/7} \left(\frac{\nu}{\rm GHz}\right)^{2/7}
\left(\frac{S_{\nu}}{\rm mJy}\right)^{4/7} \left(\frac{d}{\rm
  kpc}\right)^{8/7} \phantom{00} {\rm erg} \phantom{0} {\rm s^{-1}}
\end{equation}

The corresponding $\sim$equipartition magnetic field can therefore be estimated as

\begin{equation}
B_{\rm eq} \sim 30 \eta^{2/7} \left(\frac{S_{\nu}}{\rm mJy}\right)^{2/7} \left(\frac{d}{\rm kpc}\right)^{4/7} \left(\frac{\Delta t}{\rm s}\right)^{-6/7} \left(\frac{\nu}{\rm GHz}\right)^{1/7} \phantom{00} {\rm G}
\end{equation}

and finally, the Lorentz factors of electrons (or positrons) emitting synchrotron
emission at a given frequency can be estimated by:

\begin{equation}
\gamma_e \sim 30 \left(\frac{\nu}{\rm GHz}\right)^{1/2} \left(\frac{B}{\rm G}\right)^{-1/2}
\end{equation}

It is therefore rather straightforward to estimate a minimum energy and power associated with a given ejection event. As a strong caveat, it should be noted that the injection timescale could be much shorter than the rise time, which in many models for synchrotron emission is dominated by the evolution from large to small optical depth in the expanding plasmon, which -- in the event that all the other assumption were correct -- means that the above expression for the power is lower limit.

Fig {\ref{cygx3flare}} shows a relatively `clean' radio flare event from the X-ray
binary jet source Cyg X-3. The observation is at 15 GHz, has a rise
time of $\sim 3500$ s, an amplitude of $\sim 200$ mJy and Cyg X-3 lies
at an estimated distance of $\sim 8$ kpc. Using the above
approximations we find a minimum energy associated with the event of
$E_{\rm min} \sim 5\times 10^{40}$ erg, and a corresponding mean jet
power during the event of $\sim 10^{37}$ erg s$^{-1}$, many orders of
magnitude greater than the observed radio luminosity.
The corresponding equipartition field can be estimated
as $\sim 0.5$ Gauss, in which field electrons radiating at 15 GHz must
have Lorentz factors $\gamma \sim 150$. These solutions are illustrated in Fig \ref{minE}.

\begin{figure}
\includegraphics[scale=.65]{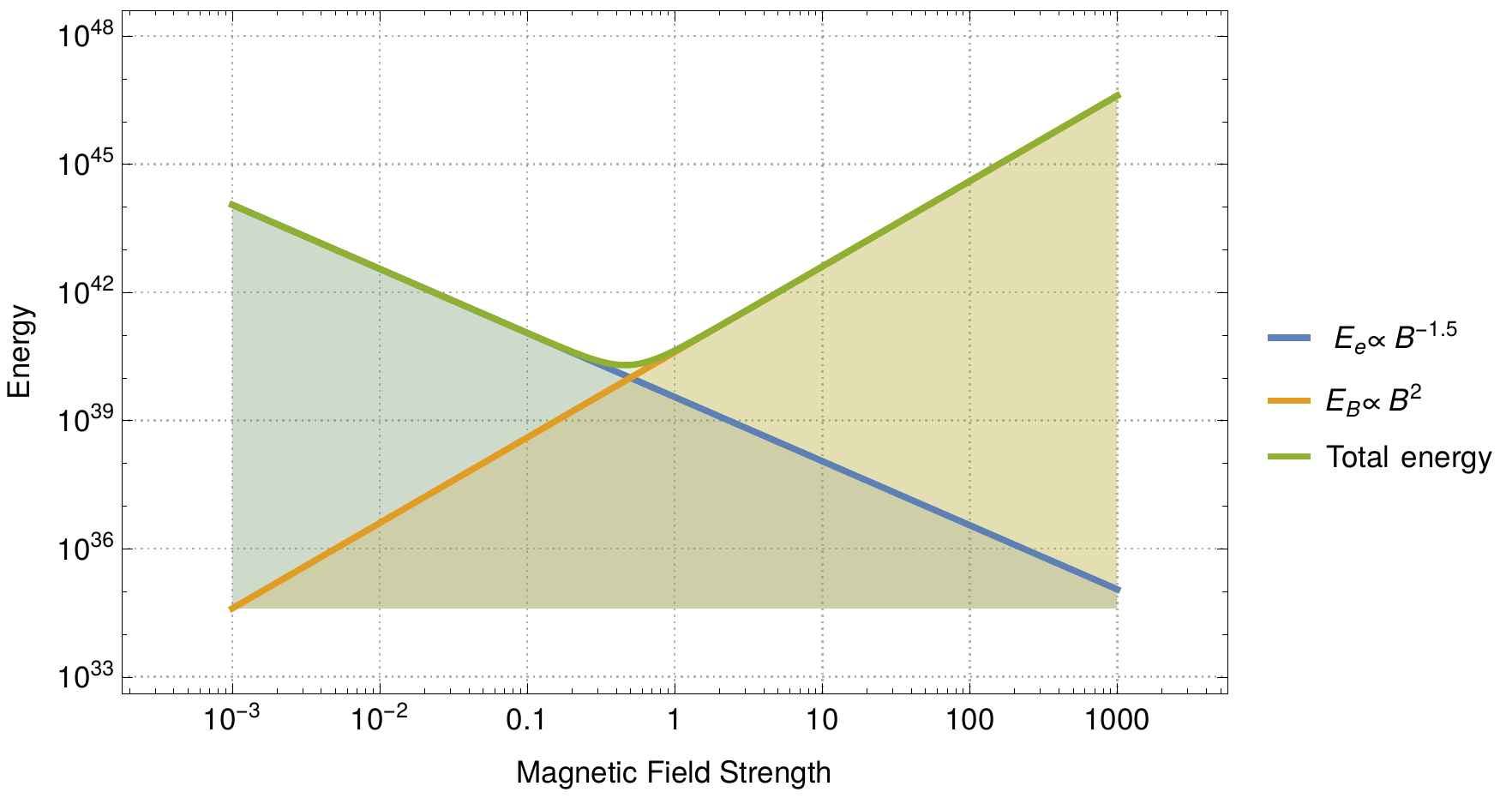}
\caption{The energy in electrons and magnetic fields for the flare illustrated in Fig \ref{cygx3flare}, using the equations outlined in the text. The minimum energy solution occurs around equipartition of energy between electrons and fields, which in this case corresponds to a magnetic field of $\sim 0.5$G.
}
\label{minE}
\end{figure}

But we're not finished yet. As already mentioned, it has been established observationally that ejections associated with such flares have bulk Lorentz factors $\Gamma \geq 2$, which means we have to both transform frequencies and timescales back to the restframe of the ejecta, and consider the kinetic energy of bulk motion as well as the internal energy in electrons and magnetic field. The kinetic energy is given by:

\begin{equation}
E_{\rm kin} = (\Gamma-1) E_{\rm int}
\end{equation}

 -- i.e. for a bulk Lorentz factor $\Gamma >2$ (by no means unreasonable --
see below) kinetic dominates over internal energy. Finally, we may also consider the additional kinetic energy which is associated with the bulk motion of one cold proton for every electron, adding yet more power. However, (i) it could be that the electrons are neutralised by positrons, not protons, (ii) the total mass of protons depends strongly on the lowest energy to which the electron spectrum extends, which is not well measured (and is somewhere that the new low-frequency radio arrays such as LOFAR and MWA can contribute). A set of calculations, with and without bulk relativistic motion and protons, is presented in Fender \& Pooley (2000), and summarised in Fig \ref{fp2000} and Table \ref{fp2000t}.

\begin{figure}
\includegraphics[scale=.5]{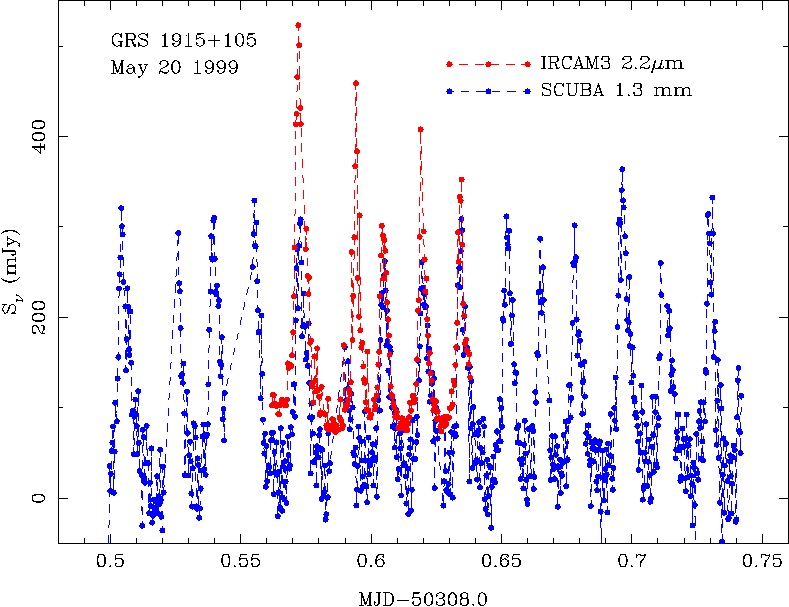}
\caption{Large, repeated synchrotron flares observed at mm (blue) and infrared (red) wavelengths from the black hole binary GRS 1915+105. The figure shows a sequence of 16 quasi-periodic flaring events observed at mm wavelengths, five of which are simultaneously detected in the infrared band. Table \ref{fp2000t} presents the minimum power calculations for these events under a number of assumption. Both are from Fender \& Pooley (2000).}
\label{fp2000}
\end{figure}

\begin{table}
\caption{
Calculation of radiative luminosity, equipartition magnetic
field, total energy and jet power \& mass-flow rate for the IR/mm events presented in Fig \ref{fp2000}, given different physical assumptions. $\Gamma$ is the
bulk motion Lorentz factor, $f$ is the `filling factor'.
In these
calculations a distance of 11 kpc and
Doppler factors for relativistic bulk motion which are the same as
those reported in Fender et al. (1999) are all assumed. Mass flow rate 
$\dot{M}_{\rm jet}$ and jet power P are
based upon one ejection every 20 min. Reproduced from Fender \& Pooley (2000).}
\label{fp2000t}
\begin{tabular}{p{1.8cm}p{1.1cm}p{1.4cm}p{1.4cm}p{1.1cm}p{1.1cm}|p{1.4cm}p{1.4cm}}
\hline\noalign{\smallskip}
Case & f & L(erg) & B$_{\rm eq}$(G) & $E_{\rm min}$(erg) & M (g) & P (erg s$^{-1})$ & $\dot{M}_{\rm jet}$(g s$^{-1}$) \\ 
\noalign{\smallskip}\svhline\noalign{\smallskip}
e$^+$:e$^-$, $\Gamma=1$ & 0.01
& $3 \times 10^{37}$ & 145 & $6 \times 10^{40}$ & -- & $5 \times
10^{37}$ & -- \\
e$^+$:e$^-$, $\Gamma=1$ & 0.1
& $3 \times 10^{37}$ & 75 & $2 \times 10^{41}$ & -- & $2 \times
10^{38}$ & -- \\
e$^+$:e$^-$, $\Gamma=1$ & 1.0
& $3 \times 10^{37}$ & 40 & $4 \times 10^{41}$ & -- & $3 \times
10^{38}$ & -- \\
e$^+$:e$^-$, $\Gamma=5$ & 1.0
& $4 \times 10^{39}$ & 115& $3 \times 10^{43}$ & -- &
$3 \times 10^{40}$ & -- \\
p$^+$:e$^{-}$, $\Gamma=1$ & 1.0
& $3 \times 10^{37}$ & 40 & $4 \times 10^{41}$ & $2 \times 10^{23}$ &
$3 \times 10^{38}$ & $2 \times 10^{20}$ \\
p$^+$:e$^{-}$, $\Gamma=5$ & 1.0
& $5 \times 10^{39}$ & 115& $1 \times 10^{46}$ & $3 \times 10^{24}$ &
$8 \times 10^{42}$ & $4 \times 10^{21}$ \\
\noalign{\smallskip}\hline\noalign{\smallskip}
\end{tabular}
\end{table}

For a more recent discussion providing an even fuller and probably more realistic treatment, the interested reader is directed to Zdziarksi et al. (2014a,b). We further note that in reality, while these discrete major ejections are fascinating and powerful events in their own right, their short duration relative to an entire outburst means that their overall contribution is probably less than the intergrated kinectic feedback from the less spectacular ($\sim$ steady) hard state jets (see Section 4).

\subsubsection{No jets}

In the soft state we do not observe core radio jets at all, implying that the radio luminosity is at least two orders of magnitude (the best limits to date, from Russell et al. 2011) below that observed at the same X-ray luminosity in the hard state. 

Using equation \ref{steadypower}, in which $L_{\rm jet} \propto L_{radio}^{12/17}$ we can estimate that this corresponds a decrease in jet power of at least $100^{12/17} \sim 25$. Of course, alternative explanations, rather than a simple decrease in core jet power, may be responsible for the decrease in the core radio flux but this remains, in our opinion, the most likely explanation (see also discussion in Fender, Homan \& Belloni 2009). 

\subsection{Winds}

The existence of accretion disc winds in black hole X-ray binaries is revealed via X-ray spectroscopy (e.g. Lee et al. 2002). They are identified by absorption features of mainly Fe {\sc xxv} and Fe {\sc xxvi} blueshifted by $\sim 1000$ km s$^{-1}$ (Fig. \ref{wind}). Neilsen \& Lee (2009) showed that in the powerful jet source GRS 1915+105, an incredibly rich source of data on accretion and jet processes, the wind was strongly and rapidly coupled to the X-ray state, being only present when the source was in (relatively) soft states (Fig. \ref{wind}). Ponti et al. (2012) studied this wind -- X-ray state coupling in a limited, but significant sample of black hole binaries, finding ubiquitous wind-tracer absorption lines during the soft states of systems viewed at high inclination, whilst they are not observed in either the hard states of the same sources and lower inclination objects. The latter implies an equatorial geometry for the wind (see also D\'iaz Trigo et al. 2006).

\begin{figure}
\includegraphics[scale=.30]{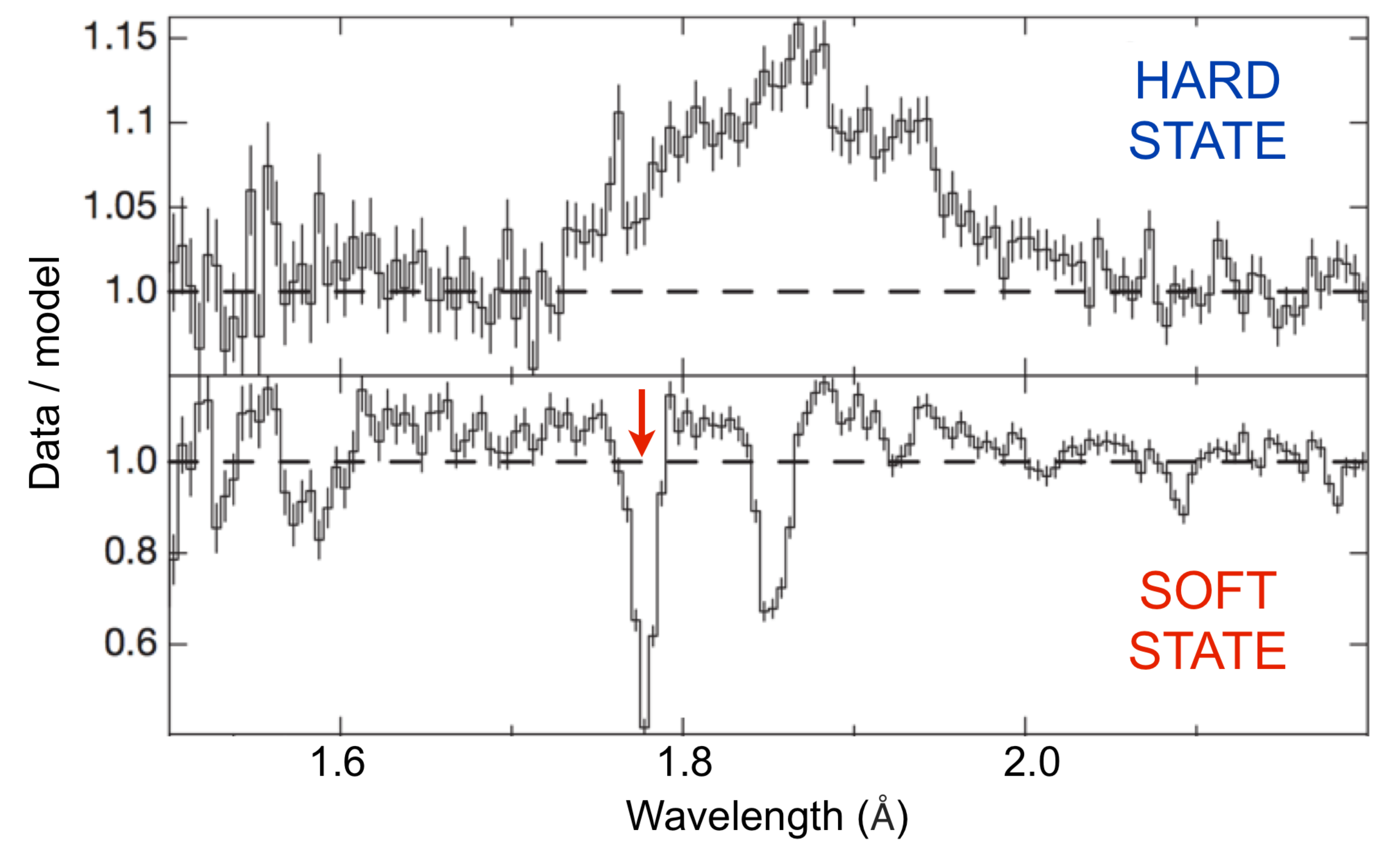}
\caption{Accretion wind -- X-ray state coupling for the case of GRS 1915+105. Several absorption lines are present during the soft state (middle and bottom panels), including a strong, blueshifted, Fe {\sc xxvi} feature (red arrow). They are absent during the hard state (top panel), where a broad Fe {\sc xxv} emission line is seen in this case. Adapted from Neilsen \& Lee 2012 (see also Lee et al. 2002).}
\label{wind}
\end{figure}

As detailed in Lee et al. (2002), simple calculations suggest that the wind might play a key role in the accretion process. From the equivalent width ($W_{\lambda}$) of the resonant absorption lines of Fe {\sc xxv} and Fe {\sc xxvi} one can estimate the column density $N_{j}$ of every species using the relation:
\begin{equation}
\frac{W_{\lambda}}{\lambda} = 8.85 \times 10^{-13} \lambda f_{ij} N_{j} 
\end{equation}
where $f_{ij}$ is the oscillator strength of the corresponding transition and $\lambda$ the wavelength (in centimetres). This relation assumes that the lines are not saturated, but in the linear part of the curve growth. The obtained ionization fraction ($N_{\rm Fe~{XXVI}} : N_{\rm Fe~XXV}$) can be compared with numerical simulations (Kallman \& Bautista 2001) in order to infer the ionization parameter ($\xi$), which defines the state of an optically thin gas:
\begin{equation}
\xi= L/nr^{2} 
\label{ip}
\end{equation}
where L is the luminosity of the incident radiation, n is the gas density, and r is the distance from the irradiating source (Tarter, Tucker, \& Salpeter 1969). In many cases, Fe {\sc xxv} and Fe {\sc xxvi} absorption lines are the only iron features in the spectrum (i.e. Hydrogen-like and Helium-like iron), which is a signature of  a highly ionized plasma. Indeed, ionization parameters of  $\xi\approx 10^4$ are typically obtained -- corresponding to temperatures above $\sim 10^6$ K -- through the above method.  From $\xi$, we can estimate the mass carried away by the wind by simply using:
\begin{equation}
\dot{M}_{\rm wind} = 4\pi r^2 n m_p v_{\rm wind} (\frac{\Omega}{4\pi})
\end{equation}
which can be re-written (using Eq. \ref{ip}) as:

\begin{equation}
\dot{M}_{\rm  wind} = 4\pi m_p v_{\rm wind} (\frac{L}{\xi})(\frac{\Omega}{4\pi})
\end{equation}
being $m_p$ the proton mass, $v_{\rm wind}$ the outflow velocity and $\Omega$ the solid angle subtended by the wind (i.e.  $\frac{\Omega}{4\pi}$ is the wind covering factor). Typical values for the wind velocity  ($v_{\rm wind} \sim 1000$~km s$^{-1}$) and the covering factor (opening angle $\sim 30^{\circ}$; Ponti et al. 2012) yield $\dot{M}_{\rm wind} \sim 10^{19}$ g s$^{-1}$, comparable to, if not larger, than the central mass accretion rate inferred from the observed luminosity. Therefore, the corresponding kinetic power carried by the wind would be of the order of $L_{\rm wind}\sim 10^{35}$ erg s$^{-1}$, significantly lower than the luminosity radiated (see section 4 below). 

\begin{figure}[t]
\includegraphics[scale=.40]{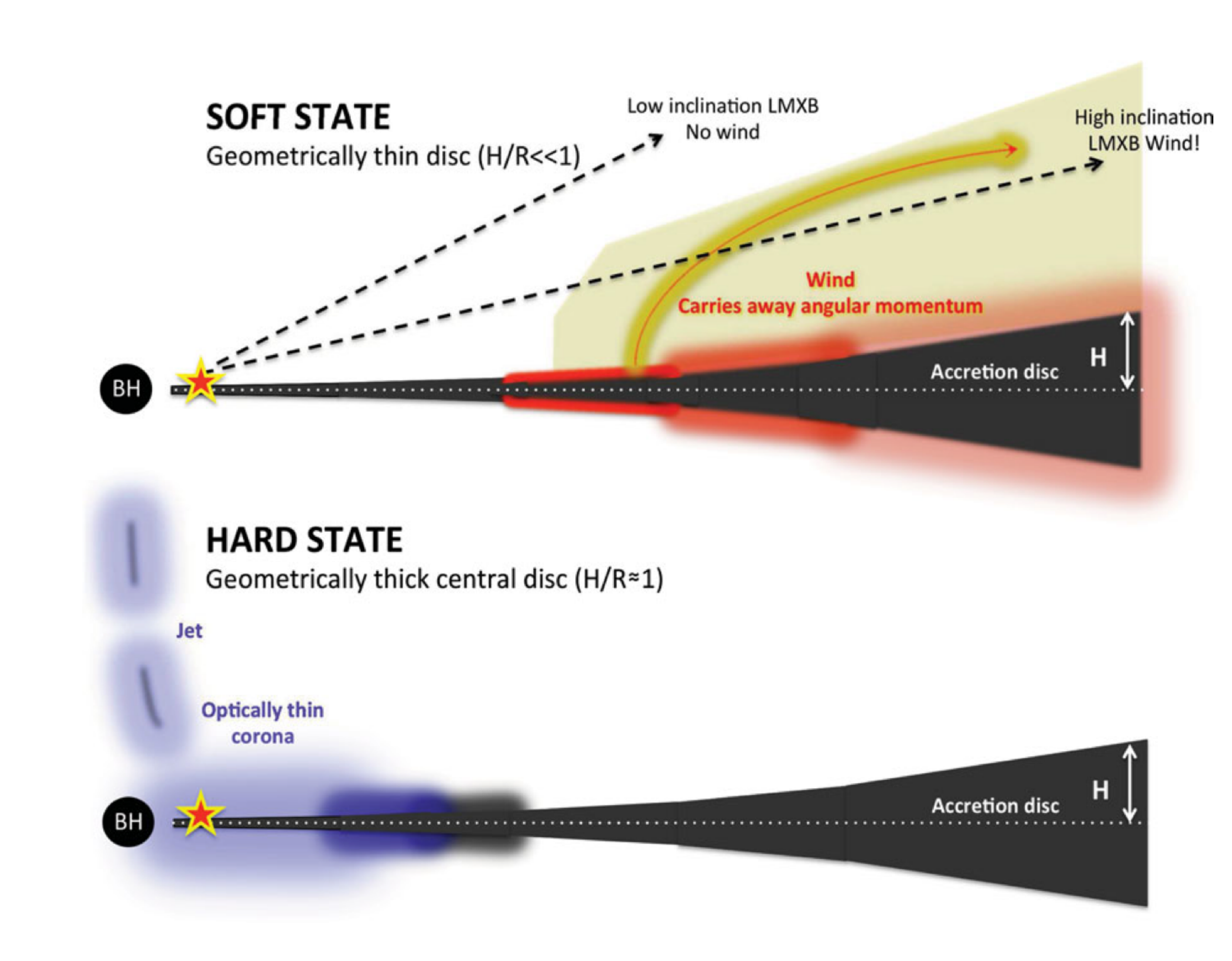}
\caption{Sketch representing the thermal wind scenario for the soft and the hard state. In the latter case, no wind would be expected if the outer disc is not sufficiently irradiated (heated). From Ponti et al. 2012.}
\label{thermal_winds}
\end{figure}

Wind outflows more than an order of magnitude larger than the contemporaneous central mass accretion rates have been estimated in GRS 1915+105 (at luminosities close to the Eddington limit) by Neilsen, Remillard \& Lee (2011). At least in this case, the properties of the wind (e.g. velocity, column density) are found to be not constant with time in response to just minor changes in the ionizing luminosity. 

Soft-state luminosities are generally higher than that of the hard-state; thus, ionization effects might play a role in explaining the observed phenomenology. To test this scenario, simulations assuming power-law irradiation with a photon index of 2 are typically used. They conclude that the observed change in luminosity is not enough to explain the large variation in ionization state required to generate/quench wind-tracer absorption lines if the same absorber material were always present (e.g. Ponti et al. 2012). However, a more accurate description of the (different) spectral energy distribution (SED) during the hard and the soft state is required to further confirm this point. This has been done for one neutron star systems that seem to show a similar wind-state coupling (Ponti et al. 2015; see also section \ref{NS_outflows}) finding consistent results. Similarly, not using self-consistent SEDs when determining the ionization parameter is also one of the main uncertainties involved in the mass outflow rate calculation described above. 

The nature of the accretion wind in X-ray binaries is still unknown and more than one mechanism (i.e. type of wind) might be at play. An appealing physical interpretation is provided by the thermal wind scenario (Begelman et al. 1983), which is sketched in Fig. \ref{wind}. Here, the outflow is produced as a consequence of the strong irradiation suffered by the outer disc, which is heated, increasing the thermal pressure to the point that a wind is driven off. Such a strong irradiation might be suppressed during the hard state, when a geometrically thick inner disc and a jet are present. Indirect evidence for the disc being less irradiated in the hard state have been recently presented in Plant et al. (2014) by studying the evolution of the ratio between Comptonized and reflected emission across the different X-ray states. 

Numerical 2D simulations performed by Luketic et al. (2010) give support to thermal winds carrying more mass than that being accreted (up to a factor of $\sim 7$), and having a strong angular dependence. However, in some cases other launching mechanisms may play a role. These might be related to the accretion disc magnetic field (i.e. magnetic winds; see e.g. Miller et al. 2006) and would imply that the amount of mass carried by the wind could be even larger.  

We will see below that, despite this large mass flow, kinetic feedback from the wind is probably not very important in the bigger picture. However, the removal of mass from the disc could potentially have a profound effect on the evolution of outbursts: in the thermal wind scenario sketched above, the duration of the soft state could be strongly limited by mass-loss in the wind, with only the mass within the wind-launching radius at the time of the state transition able to ultimately make it to the black hole.

\section{Feedback throughout an outburst}

Given the relations outlined in the previous section, we are now in a position to take radio and X-ray measurements made throughout the outburst of a black hole X-ray binary and use them to build up a picture of how feedback proceeds and accumulates. In the following, we use an outburst of the black hole binary GX 339-4, which took place in 2002, and was well monitored in the X-ray band. In order to smoothly accumulate our functions through the outburst, we used the smoothed, multiple Gaussian fit to the outburst also used in Coriat, Fender \& Dubus (2012). Fig \ref{Light} presents this light curve, broadly separated into hard and soft states based upon the timing properties as observed by RXTE (see Belloni, Motta \& Mu\~noz-Darias 2011).

\begin{figure}
\includegraphics[scale=.5]{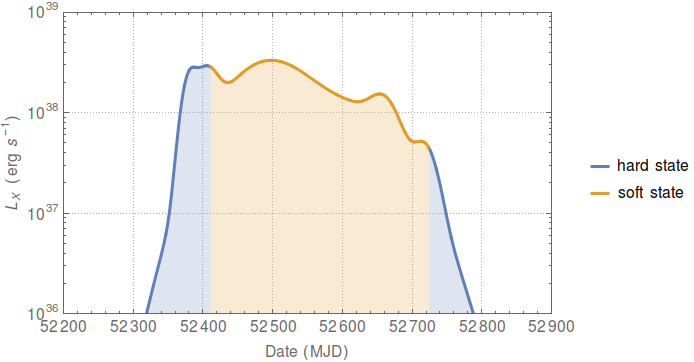}
\caption{A smoothed fit to the X-ray lightcurve of the 2002-2003 outburst of the black hole X-ray binary GX 339-4, roughly separated into hard and soft X-ray states based on the timing properties. The curves are based on functions fit in Coriat, Fender \& Dubus (2012), and correspond to the flux in the 0.1--200 keV energy range.}
\label{Light}
\end{figure}

\begin{figure}
\includegraphics[scale=.5]{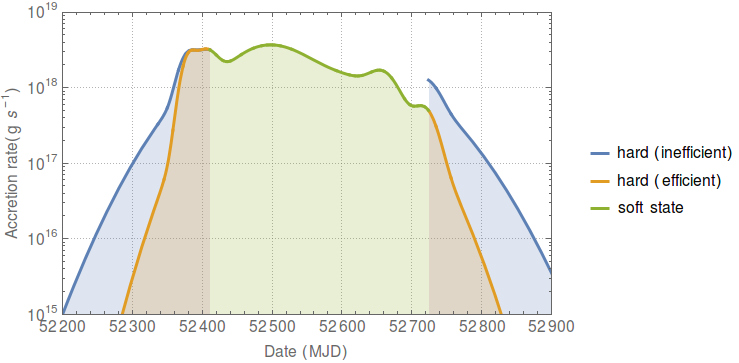}
\caption{The accretion rate estimated as outlined in the text, for the outburst of GX 339-4. The accretion rate during the soft state is converted linearly from the X-ray flux, assuming radiatively efficient ($\eta = 0.1$) accretion and no significant bolometric correction. During the hard state the accretion rate is measured via two different approaches. The orange light curve corresponds to radiatively efficient hard states, using the same prescription as for the soft state. The blue light curve corresponds to radiatively inefficient accretion, where $\eta =0.1$ at the hard $\rightarrow$ soft state transition, but then falls linearly with $\dot{m}$ thereafter (corresponding to $L_X \propto \dot{m}^2$). This approach, which is widely accepted and messes nicely with models for accretion and jet formation, results in a strong discontinuity at the soft $\rightarrow$ hard state transition, resulting from the fact that the radiative efficiency should drop significantly, and yet no strong step down is observed in the X-ray light curve. The resolution of this issue is not yet clear, but the two hard state curves probably encompass the reasonable range of possibilities.}
\label{Mass}
\end{figure}

\begin{figure}
\includegraphics[scale=.5]{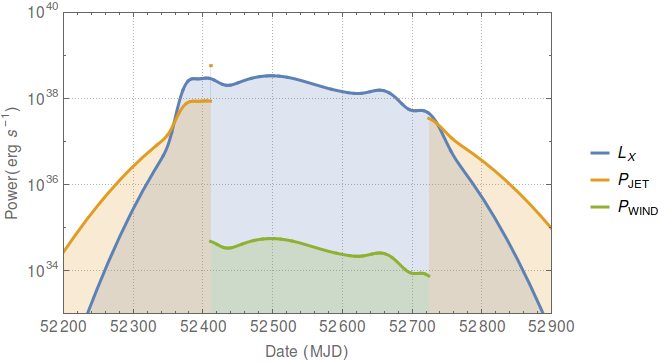}
\caption{A comparison of the observed X-ray radiative luminosity with the kinetic powers produced by the jet (hard state and state transition) and the wind (soft state). The kinetic power of the wind is extremely small compared to both the radiation and jet kinetic power at their peaks; however it may still play an important role in regulating the evolution of outbursts. Away from the peak of the outburst, the source is predicted to enter `jet dominated' states where the jet kinetic power exceeds the radiative luminosity (Fender, Gallo \& Jonker 2003).}
\label{Power}
\end{figure}

We take the methods discussed in the previous sections, and combine them with the following assumptions:

\begin{enumerate}
\item{In the soft state the radiative efficiency $\eta$ is 0.1, which allows us to calculate the accretion rate directly.}

\item{At the hard $\rightarrow$ soft state transition, the radiative efficiency of the hard state is also $\eta = 0.1$, ensuring no strong discontinuity in the light curve (but see below), but either remains constant in the hard state ({\em radiatively efficient hard state}) or falls linearly with with accretion rate below that in the hard state, such that $L_{X, hard} \propto \dot{m}$ ({\em radiatively inefficient hard state}). This allows us to calculate the accretion rate in the hard state.}
\item{We further assume that at the hard $\rightarrow$ soft state transition, the core jet kinetic power is 10\% of the radiative output, and falls linearly with the accretion rate, as implied by the observations and the scalings discussed earlier. This is consistent with applying equation [7] to the empirical radio:X-ray correlation for GX 339-4.}
\item{During the soft state, we assume that three times as much mass is being lost in the accretion disc wind as is being accreted centrally.}
\item{Finally, we assume that following the state transition there is a period of major relativistic ejections, which lasts one day and during which period the jet power is twice the observed X-ray luminosity (or 200 times more powerful than the peak hard state jet, which is at the upper end of power estimates).}
\end{enumerate}

In Fig \ref{Mass} we plot the mass accretion rate, as inferred from the X-ray light curve, under the assumptions of both radiatively efficient and radiatively inefficient accretion in the hard X-ray state. We see that, as mentioned earlier, in the case of a simple switch from $L_X \propto \dot{m}$ to $L_X \propto \dot{m}^2$ at the soft $\rightarrow$ transition, there is a strong discontinuity in the inferred mass flow. 

Fig \ref{Power} plots our estimates of the total power in radiation, jets and winds throughout the outburst. The early and final stages of the outburst see the source in a jet-dominated state (Fender, Gallo \& Jonker 2003), transitioning to radiation-dominated at or shortly before the state change. A brief period of radio flaring during the state change {\em may} exceed the radiation in power, but this is for a very short period and is probably inconsequential in terms of the integrated feedback. The kinetic feedback from the wind, even with a very large mass outflow rate, is orders of magnitude below the peak radiation and jet powers, and is insignificant for the feedback of kinetic energy to the ambient environment.

\begin{figure}
\includegraphics[scale=.65]{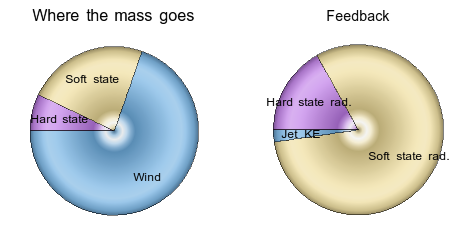}
\caption{Mass flow and feedback estimated for the outburst of GX 339-4 using the assumptions stated in the text. The {\bf left panel} indicates where the inflowing mass ended up; the majority was lost in the accretion disc wind, and most of the central accretion occurred during the soft state. Note that this figure is for the case of radiatively inefficient accretion in the hard state ($c=2$) but the results are not very different for efficient hard state accretion. The {\bf right panel} summarises the
radiative and kinetic feedback integrated over the course of the outburst. Radiation, the dominant fraction of which arises in the soft X-ray state, dominates over kinetic feedback from the jet. The kinetic feedback from the wind is completely insignificant and is not plotted here.
}
\label{powerpie}
\end{figure}

So what's the integrated feedback? Fig \ref{powerpie} attempts to illustrate this by comparing the total energy in hard and soft state radiation, and in kinetic feedback in the jet. Before taking these results at face value, recall however, that:

\begin{itemize}
\item{The kinetic power estimates, especially for the jet, are very uncertain (as you should have learned from reading this review).}
\item{This integrated power does not include the {\em jet-dominated} feedback during long periods (years) of quiescence (although it seems likely that feedback is dominated by the outbursts, at least for this type of system and behaviour).}
\item{It may well be that it is far easier, per erg of feedback, for the jets to significantly affect their environments than the radiation (obviously the case if the environment is entirely optically thin).}
\end{itemize}

\subsection{Where does this leave the energy balance?}

Earlier in the review, we discussed how accurate measurement of the feedback terms in equation [4] could potentially lead to estimates of the contributions from black hole spin and advection. However (as forewarned), we see that in the analysis in this section, we are forced to use the observed radiative luminosity to estimate the accretion rate, thereby already relegating the spin and advection terms to (assumed) minor significance. It is entirely possible (although theoretically unlikely), that the $L_{advected}$ term is an order of magnitude larger than all the other terms in equation [4], but we are simply unaware of this due to the (quasi-circular) assumptions made in the above analysis. So the promise of measuring these important terms turns out to have been a bit of an illusion, for now. How might we, in future, try to test these terms?

The rate of mass accretion can, in principle, be estimated from the binary parameters of a system (e.g. Coriat et al. 2012 and references therein). Those estimates seem to support the assumption that, during the phases when most of the matter is accreted, it is done so in a radiatively efficient way (i.e. $L_{advected}$ is a minor term). This conclusion is supported, completely independently, by the `Soltan argument' where the cosmic X-ray background (a record of accretion onto AGN, peaking $1 \leq z \leq 2$) can be compared to the local space density of supermassive black holes (e.g. Soltan 1982; Fabian 2012 and references therein).

What about the spin term? At present this also seems likely to be small contributor to the overall energy budget of outbursts. However, its role in producing powerful jets has long been advocated for AGN (see e.g. Sikora, Stawarz \& Lasota 2007) or even just assumed (e.g. Ghisellini et al. 2014). For X-ray binaries, in recent years `direct' spin measurements have emerged via two methods of X-ray spectroscopy (e.g.  McClintock, Narayan \& Steiner 2014; Miller 2007; Reynolds 2014) and also, most recently, X-ray timing (Motta et al. 2014a,b). It is not at all clear that the jet power estimates, measured from the radio emission, correlate in any way with these reported spins. This conclusion is strongest for the hard state sources, which appear to dominate the overall kinetic feedback during an outburst (Fender, Gallo \& Russell 2010 and the analysis in the previous section) but has been challenged in the case of the transient state transition flares (Narayan \& McClintock 2010; the interested reader can follow the debate in Steiner, McClintock \& Narayan 2013; Russell, Gallo \& Fender 2013; Middleton et al. 2013 and citations thereof).

\section{Applicability to AGN}

Does any of this have anything to do with supermassive black holes in AGN, the feedback from which is thought to regulate the growth of galaxies and keep the gas in the centres of clusters hot (e.g. McNamara \& Nulsen 2007)? There are some reasons to believe so but, again, this is not proven. 

\subsection{Power and accretion across the mass spectrum}

\begin{figure}
\includegraphics[scale=.675]{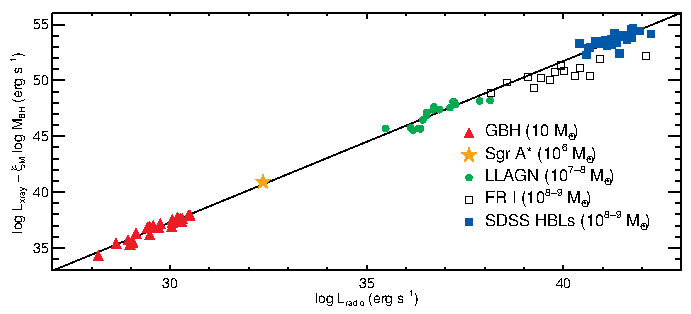}
\caption{The `fundamental plane of black hole activity' demonstrating a clear correlation between (core) radio and X-ray luminosities across all black holes when a mass term is considered. This representation is from Plotkin et al. (2012), and considers only relatively low accretion rate sources (see also Falcke, K\"ording \& Markoff 2004; K\"ording, Falcke \& Corbel 2006), but the relation also holds, albeit with more scatter, when all types of AGN are included (Merloni, Heinz \& di Matteo 2003). All of the aformentioned papers present versions of this fundamental plane; the representation here is shown to be further improved by making corrections for Doppler boosting in some of the AGN (see Plotkin et al. 2012 for all the details).
The acronyms are as follows: GBH = Galactic Black Holes (BHXRB in this review), Sgr A* = Sagittarius A* (galactic centre black hole), LLAGN = Low luminosity AGN, FR I = Fanaroff-Riley I radio sources, SDSS HBL = Sloan Digital Sky Survey High-Energy cut off BL Lac obkjects, and $\zeta \sim -0.9$. Note that for the X-ray binaries, both the `intrinsic' and `global' effects are seen, in the sense that individual sources are observed to move up and down along the overall population (see Fig \ref{coriat}), formed of multiple sources. To date, the AGN correlation is `global' only, formed by the population.}
\label{fp2}
\end{figure}

Following the discovery of the `universal' radio:X-ray correlation (Gallo, Fender \& Pooley 2003; Corbel et al. 2012) in X-ray binaries, Merloni, Heinz \& di Matteo (2003) and Falcke, K\"ording \& Markoff (2004) rapidly established a relation which also encompassed AGN when a mass term was considered (and only core, not extended, radio emission was used). This joint discovery of a `fundamental' plane marked a milestone in connecting feedback from black holes of all masses. In a sense this extended the steady-jet, hard state correlation, from stellar mass to supermassive black holes. Fig \ref{fp2} presents the fundamental plane in the representation of Merloni et al., with the addition of the detection of radio emission from the nearby stellar mass black hole A0620-00 in quiescence (Gallo et al. 2006), demonstrating the applicability of the relation from the weakest to the most powerful black hole jets. As already noted, the functions used to approximate the relation between core radio luminosity and intrinsic jet power are also very similar, consistent with being identical. 

\subsection{The disc-jet coupling}

\begin{figure}
\includegraphics[scale=.4]{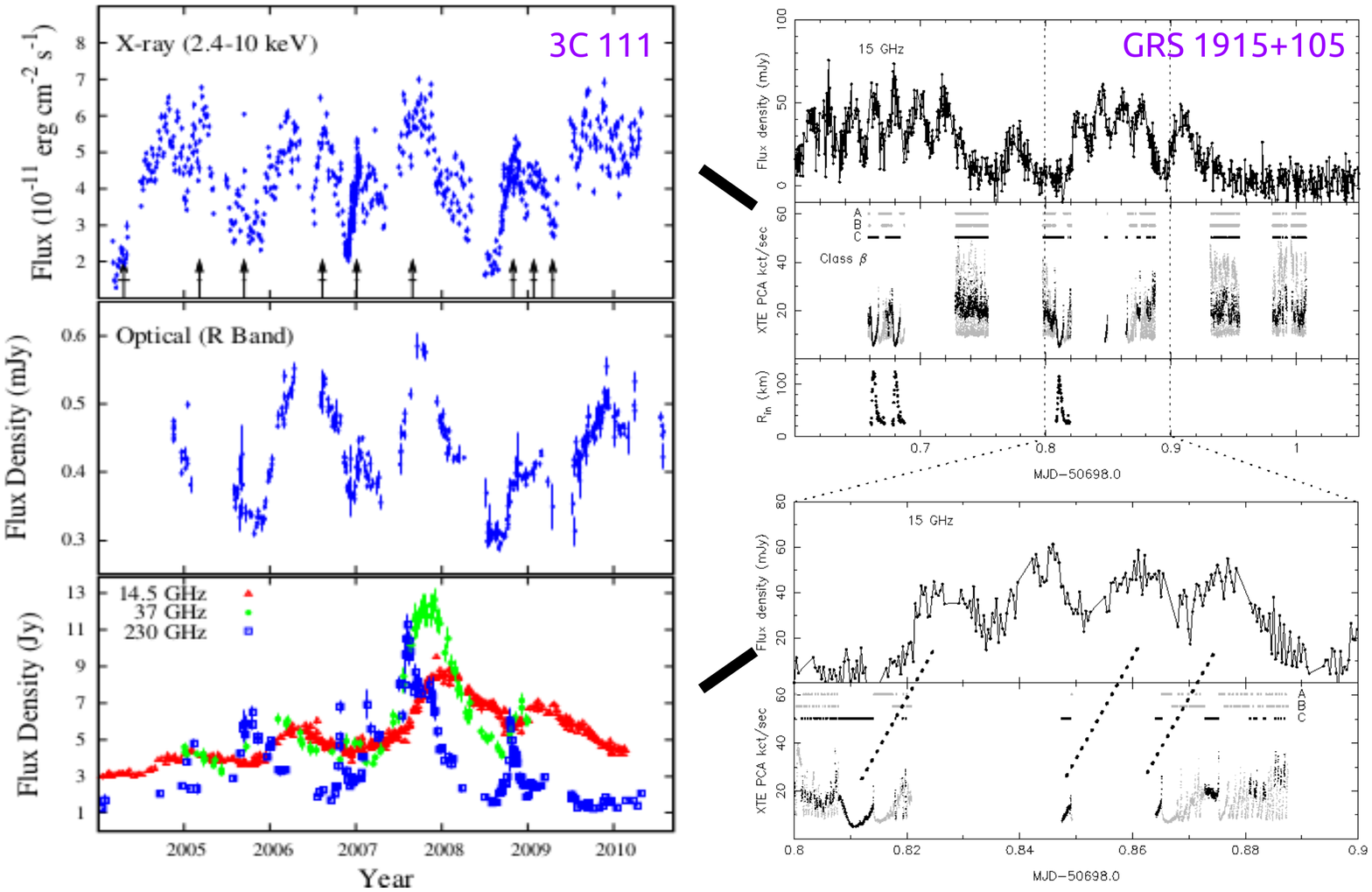}
\caption{Qualitatively similar accretion:ejection coupling in an AGN and an X-ray binary.
In the {\bf left panel}, the connection between superluminal ejections, radio and optical flaring, and X-ray dips in the radio galaxy 3C 111 (Chatterjee et al. 2011). In the {\ right panel}, the connection between radio flares, which we are confident correspond to ejection events, and X-ray dips, in the X-ray binary GRS 1915+105 (Klein-Wolt et al. 2002). In this source we also see that periods without strong X-ray dips do not produce radio flares, and indeed there is a remarkable, complex and poorly-understood phenomenology of radio:X-ray connection in this system (see e.g. review in Fender \& Belloni 2004).}
\label{dips}
\end{figure}

In K\"ording, Jester \& Fender (2006) it was argued that, qualitatively, the relation between whether or not an AGN was radio loud and its accretion rate and broadband spectrum was analagous to that observed in stellar mass black holes. Most convincingly, it seems that AGN at relatively low accretion rates (LLAGN) are relatively radio loud (see also e.g. Ho 2002), like the hard state of black hole X-ray binaries. This connects the states (at high Eddington ratios) of stellar mass black holes to AGN and is also of great interest, not least because it might well be enough on its own to explain why only $\sim 10$\% of luminous AGN are radio-loud (e.g. discussion in Sikora, Stawarz \& Lasota 2007), since this is approximately the fraction of time spent in radio-loud states during an outburst (as we've seen in Section 4, above). In addition to these population-based comparisons, individual AGN have shown behaviour which appears to be directly comparable to the accretion-outflow coupling in the black hole binary GRS 1915+105. Marscher et al. (2002; see also Chatterjee et al. 2009; 2011) have reported spatially-resolved ejecta from the blazars 3C 120 and 3C 111 which appear to be associated with X-ray dips from the central AGN. Recent work by Lohfink et al. (2013) appears to support this interpretation. Fig \ref{dips} compare the dipping/flaring behaviour in 3C 111 with that observed in the binary GRS 1915+105.

In the case of winds, there is a richer phenomenology and probably broader range of underlying astrophysics, in the case of AGN. The high ionisation in the inner parts of the accretion flow of an X-ray binary mean that fast UV-driven winds, manifest as the broad line region (probably) in AGN, are not present in the stellar mass black holes. There does appear to be a broad anti-correlation between the strongest jet sources and the presence of winds in AGN, but it is too early to claim that this is analogous to what we see in X-ray binaries.

Finally, independent of the connection to jets, there are multiple pieces of evidence that black hole accretion can be scaled across the vast mass range from stellar to the most supermassive, for example in the relation between black hole mass, accretion rate and timing properties (e.g. McHardy et al. 2006). It is undoubtedly the case that the linked fields of BHXRB and AGN accretion can learn much from each other, and we would encourage each community to (occasionally) review the status of the other.

\section{Comparison to neutron stars -- the most important control sample}

In the local Universe, neutron stars are more common than stellar mass black holes (factor of $\sim 10 $), and so are neutron stars in X-ray binaries. They totally dominate the population of persistent systems (i.e. those always active above $\sim$ 1\% $L_\mathrm{Edd}$), and significantly contribute to the transients, the group where the vast majority of black holes are found. It is therefore not surprising that, on a given day, nine of the ten brightest sources in the X-ray sky are accreting neutron stars. These have yielded a wealth of observations that can be equally used to study accretion processes in the strong gravitational field regime (see van der Klis 2006 for a review). 

However, there are, at least, two basic differences to bear in mind. (i) Neutron star have a surface, where up to $50$ \% of the accretion luminosity might be radiated (Sunyaev \& Shakura 1986), and (ii) an anchored magnetic field. The former has a clear impact in their energy spectrum (and thus, colours) and have resulted in a variety of alternative spectral models (e.g. White, Stella \& Parmar 1988; Mitsuda et al. 1989). On the other hand, although low mass X-ray binaries harbour old, weakly magnetized neutron stars, magnetic fields might still modify the accretion geometry, as can occur in some compact binaries with white dwarf accretors. Indeed, X-ray pulsations are seen on several objects and differences in magnetic field has been one of the classical explanations for the presence of two different X-ray phenomenologies (Hasinger \& van der Klis, 1989), which are now more certainly known to be associated with different accretion rates/luminosities (e.g. Lin, Remillard \& Homan 2009). Nevertheless, anchored/central magnetic fields are likely to affect outflows properties at some level.

All things considered, it is well established that neutron stars in low mass X-ray binaries share X-ray spectral (Done \& Gierlinksi 2003; Lin, Remillard \& Homan 2007) and timing properties (Wijnands \& van der Klis 1999) with their analogous black hole systems. Recently, Mu\~noz-Darias et al. (2014) have presented the luminosity-variability plane -- firstly introduced for black holes by Mu\~noz-Darias, Motta \& Belloni 2011 as an equivalent to Fig. \ref{FB2012} -- as a common framework to study evolution of the accretion flow in both populations. In particular, the hysteresis patterns between hard and soft states, characteristic of black holes, seem to be equally common in neutron star systems accreting below $\sim 30\%$ of $L_\mathrm{Edd}$. Brighter systems stay in the soft state, and some of them -- probably the brightest -- also display \textit{soft $\rightarrow$ hard} and \textit{hard $\rightarrow$ soft} transitions, with X-ray and radio properties (see below) similar to those observed in black holes (see Fig \ref{RID}). 

\begin{figure}
\includegraphics[scale=.2]{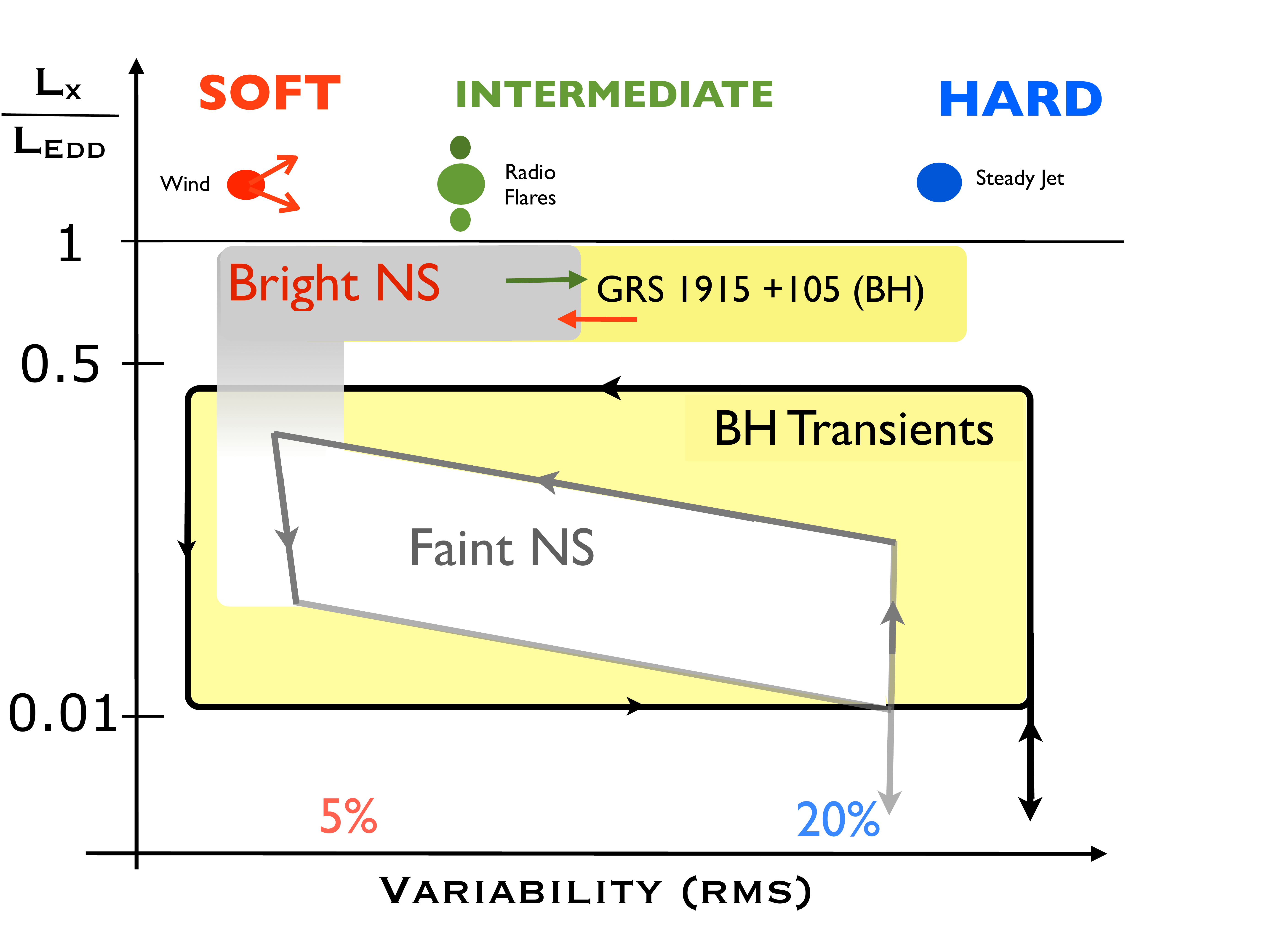}
\caption{Sketch describing the qualitative evolution and location of faint and bright neutron star (NS) systems (known as atoll and zeta sources, respectively), in the X-ray variability-luminosity plane on top of what is typically observed in black hole (BH) transients. Variability is indicated as fractional root mean square amplitude (rms) and luminosity is scaled by the Eddington luminosity. Black holes and Neutron stars are observed to display hysteresis at sub-Eddington rates. The case of GRS 1915+105, a persistent black hole that usually sits in the bright hard states is also shown. It occasionally samples softer regions, overlapping with those typical for the bright NS during their \textit{soft $\rightarrow$ hard} excursions. Both state transitions yield strong jet emission. In the upper panel we indicate the outflow properties of black holes as a function of the state, a relation that seem to hold as well for neutron stars. Adapted from Mu\~noz-Darias et al. 2014, where we direct the reader for further details.}
\label{RID}
\end{figure}

\subsection{Outflows in Neutron stars}
\label{NS_outflows}
Given the aforementioned similarities between accreting neutron stars and black holes, and the discussed accretion-ejection connexion observed for the latter group, one would expect to observe a similar outflow-X-ray state relation in the former. Indeed, radio jets, both in the form of steady outflows and discrete ejections, have been observed in these objects (see Migliari \& Fender 2006 for a global study).  Steady jets are also associated with the hard state, although for a given X-ray (accretion) luminosity they are (probably, slightly) less powerful than those in black holes. On the other hand, jet emission is only observed to quench during the bright phases of the soft state (e.g. Migliari et al. 2003), whilst radio emission seems to be present during faint soft states. As noted in Mu\~noz-Darias et al. (2014), this is in agreement with spectral studies (Lin et al. 2007) suggesting moderate Comptonization levels during this phase, and with the amount of X-ray variability.

The brightest neutron stars (known as Z sources) can be active radio sources, displaying transient jets very similar to those observed during the \textit{hard $\rightarrow$ soft} transitions of black hole systems. In some cases, such as the jets from the neutron star Cir X-1, these can be spectacular (Fender et al. 2004; Miller-Jones et al. 2012a), and we should not also forget that a couple of other powerful jet sources, e.g. Cyg X-3 and SS433, might yet turn to harbour neutron stars.
The Z sources are (generally) persistent but variable sources, and their behaviour resembles that of the (quasi-persistent{\footnote{At the time of writing, this `transient' has been active for 20 years}} but variable) black hole GRS~1915+105. As can be seen in Fig. \ref{RID}, the ejections seem to occur in a similar region of the luminosity-variability plane. We note, however, that whereas GRS~1915-105 typically sits in the bright hard state making soft excursions that trigger the radio flaring, bright neutron stars are soft state systems making hard excursions resulting in comparable radio behaviour. The reason why neutron stars are not found in bright hard states is unknown.

As it occurs with the jet outflows, the presence of winds in accreting neutron stars is also well established. They have been observed in several objects and share similar properties to those in black holes, including alike absorbers and equatorial geometries (see e.g. D\'iaz Trigo \& Boirin 2013). Recent studies (Ponti, Mu\~noz-Darias \& Fender 2014; Ponti et al. 2014) have also shown that the X-ray state--wind connection observed in black holes might also hold in these objects. In particular, these works show that deep absorption lines, like those typically associated with winds outflows, are present in the X-ray spectra of two systems during the soft state. These features are not present during the hard state of the same sources. To derive more global conclusions, the properties of these winds need to be investigated with higher resolution instruments (e.g. to measure outflow velocities), and from a larger sample of objects. However, we note that some of the systems where winds have been already detected are bright neutron stars (Z sources), and they spend most of their lives in the soft state.
Beyond neutron stars, there are hints that similar patterns of disc-jet coupling may occur in white dwarfs. K\"ording et al. (2009) reported the detection of a radio flare from the cataclysmic variable SS Cyg, which appeared to coincide with a high luminosity state transition, analogous (possibly) to that observed in black holes and neutron stars. This cataclysmic variable also displayed a hysteresis pattern that resembles those typically seen in X-ray binaries.

\section{Conclusions}

The astrophysics of accretion and space time around stellar mass black holes is extremely rich and probably our best test of classical general relativity. It is also the subject of a completely different review to this one. Neither, although we are concerned with feedback, have we considered how it works on the largest scales. 
Rather, what we have attempted to do here is to show how one could -- at least in principle -- use X-ray and radio observations to measure the total feedback, via radiation, winds and jets, associated with accretion onto these objects. By considering black holes as black box engines, and learning empirically what they do, we can try to extrapolate to understand the past and future flow of heat and energy in the universe.

In this review we have built up the background equations required to make these calculations, and have tried them out on the outburst of a black hole X-ray binary. 
We see that in terms of energy alone, our best estimates suggest that radiation dominates over kinetic feedback. Within the realm of kinetic feedback, the jet dominates the wind completely. It should be borne in mind, however, that this was for a single outburst of a single source, and -- as the reader will have appreciated throughout the earlier sections -- the estimator methods are fraught with uncertainties. Intriguingly, although the wind is entirely overshadowed by the jet in terms of integrated kinetic power, it may well dominate the mass flow and therefore regulate the duration of the outburst (and hence the overall feedback). Improved precision of the relations between radio flux and jet power, and between X-ray luminosity and accretion rate, will allow us not only to understand how much power we're going to get from our black hole engine, but also to see how much that engine is growing and being spun up or down. We remain convinced that everything we learn about cycles of accretion and feedback in stellar mass black holes has some relevance to supermassive black hole accretion, and look forward to a future of more communication and cross-fertilisation between the fields.

Finally, we have shown that with every advance in our understanding of the phenomenology of feedback in stellar mass black holes, neutron stars are keeping step. They behave extremely similarly to black holes, but also slightly differently, providing us with undoubtedly the best tests of what may or may not be unique to black holes.

\begin{acknowledgement}
RF would like to acknowledge an uncountable number of useful conversations with collaborators, friends and occasional rivals. He would also like to acknowledge the hospitality provided in Como during the school which resulted in this book, and the support of the editor, Francesco Haardt. Micka\"el Coriat kindly supplied the functional fit to the outburst of GX 339-4.
TMD would like to acknowledge the support and research opportunities provided by the EU programs \textit{Black Hole Universe} (Initial Training Network 215212) and \textit{Marie Curie Intra-European Fellowship 2011-301355}, during his research positions in INAF-Brera, Southampton and Oxford, where some of the ideas discussed here were developed. He would like also to acknowledge all the collaborators, from PhD students to heads of group, involved in this intense period of research work. This work was partially supported by ERC grant 267697 "4 PI SKY: Extreme Astrophysics with Revolutionary Radio Telescopes".
\end{acknowledgement}
\section*{Appendix}
\addcontentsline{toc}{section}{Appendix}


\begin{thebibliography}{99.}%
%
%
\bibitem{} Balbus S.A.: Enhanced Angular Momentum Transport in Accretion Disks. ARA\&A, 41, 555 (2003)

\bibitem{} Begelman M.C., Armitage P.J.: A Mechanism for Hysteresis in Black Hole Binary State Transitions. ApJ, \textbf{782}, L18 (2014)

\bibitem{}Begelman M. C., McKee C. F., Shields G. A.:   Compton heated winds and coronae above accretion disks. I Dynamics,  ApJ,  \textbf{271},  70,  (1983)

\bibitem{}Belloni T. M., Motta S. E., Mu\~noz-Darias T.:   Black hole transients,  BASI,  \textbf{39},  409,  (2011) 

\bibitem{1994ApJ...423..659B} Brown, G.~E., \& Bethe, H.~A.: A Scenario for a Large Number of Low-Mass Black Holes in the Galaxy. ApJ, \textbf{423}, 659 (1994)

\bibitem{} Burbidge G.R.: Estimates of the Total Energy in Particles and Magnetic Field in the Non-Thermal Radio Sources. ApJ, \textbf{129}, 849 (1959)

\bibitem{} Casares, J., Jonker P.G: Mass Measurements of Stellar and Intermediate-Mass Black Holes. Space Science Reviews \textbf{183}, 223 (2014)

\bibitem{} Chatterjee R. {\em et. al.}: Connection Between the Accretion Disk and Jet in the Radio Galaxy 3C 111. ApJ \textbf{734}, 43

\bibitem{} Corbel S., Nowak M.A., Fender R.P., Tzioumis A.K., Markoff S.: Radio/X-ray correlation in the low/hard state of GX 339-4. A\&A, \textbf{400}, 1007 (2003)

\bibitem{} Corbel S., Coriat M., Brocksopp C., Tzioumis A.K., Fender R.P., Tomsick J.A., Buxton M.M., Bailyn C.D.: The `universal' radio/X-ray flux correlation: the case study of the black hole GX 339-4. MNRAS, \textbf{428}, 2500 (2013)

\bibitem{} Coriat M. {\em et. al.}: Radiatively efficient accreting black holes in the hard state: the case study of H1743-322.
MNRAS, \textbf{414}, 677 (2011)

\bibitem{} Coriat M., Fender R.P., Dubus G.: Revisiting a fundamental test of the disc instability model for X-ray binaries. MNRAS, \textbf{424}, 1991 (2012)

\bibitem{2010Natur.467.1081D} Demorest, P.~B., 
Pennucci, T., Ransom, S.~M., Roberts, M.~S.~E., \& Hessels, J.~W.~T.: A two-solar-mass neutron star measured using Shapiro delay Nature, \textbf{467}, 1081 (2010)


\bibitem{}Díaz Trigo M., Boirin L.:   Disc Atmospheres and Winds in X-ray Binaries,  AcPol, \textbf{ 53},  659,  (2013)

\bibitem{}D\'iaz Trigo M., Parmar A. N., Boirin L., Méndez M., Kaastra J. S.:   Spectral changes during dipping in low-mass X-ray binaries due to highly-ionized absorbers,  A\&A,  \textbf{445},  179,  (2006) 

\bibitem{}Done C., Gierliński M.:   Observing the effects of the event horizon in black holes,  MNRAS,  \textbf{342},  1041,  (2003) 

\bibitem{} Esin A.A., McClintock J.E., Narayan R.: Advection-Dominated Accretion and the Spectral States of Black Hole X-Ray Binaries: Application to Nova Muscae 1991. ApJ, 489, 865 (1997)

\bibitem{} Fabian A. Observational Evidence of Active Galactic Nuclei Feedback. ARA\&A, \textbf{50}, 455 (2012)

\bibitem{} Fender R.P., Pooley G.G.:Giant repeated ejections from GRS 1915+105. MNRAS, \textbf{318}, L1, 2000

\bibitem{} Fender R.P., Belloni T.M., Gallo E.: Towards a unified model for black hole X-ray binary jets. MNRAS, \textbf{355}, 1105 (2004)

\bibitem{} Fender R.P., Homan J., Belloni T.M.: Jets from black hole X-ray binaries: testing, refining and extending empirical models for the coupling to X-rays. MNRAS, \textbf{396}, 1370 (2009)

\bibitem{} Fender R.P., Belloni T.M.: Stellar-Mass Black Holes and Ultraluminous X-ray Sources. Science. \textbf{337}, 540 (2012)

\bibitem{} Fender R.P., Wu K., Johnston H., Tzioumius T., Jonker P., Spencer R., van der Klis M.: An ultra-relativistic outflow from a neutron star accreting gas from a companion. Nature, \textbf{427}, 222 (2004)

\bibitem{} Fender R.P., Garrington S.T., McKay D.J., Muxlow T.W.B., Pooley G.G., Spencer R.E., Stirling A.M., Waltman E.B.: MERLIN observations of relativistic ejections from GRS 1915+105. MNRAS, \textbf{304}, 865 (1999)

\bibitem{} Feng Y., Narayan R.: Hot Accretion Flows Around Black Holes. ARA\&A, 52, 529 (2014)

\bibitem{} Frank J., King A., Raine D.J.: Accretion Power in Astrophysics: Third Edition. Cambridge University Press (2002)

\bibitem{} Gallo E., Fender R.P., Miller-Jones J.C.A., Merloni A., Jonker P.G., Heinz S., Maccarone T.J., van der Klis M.: A radio-emitting outflow in the quiescent state of A0620-00: implications for modelling low-luminosity black hole binaries. MNRAS, \textbf{370}, 1351 (2006)

\bibitem{} Harmon B.A., Wilson C.A., Zhang S.N., Paciesas W.S., Fishman G.J., Hjellming R.M., Rupen M.P., Scott D.M., Briggs M.S., Rubin B.C.: Correlations between X-ray outbursts and relativistic ejections in the X-ray transient GRO J1655 - 40. Nature, \textbf{374}, 703 (1995)

\bibitem{1989A&A...225...79H} Hasinger, G., \& van der Klis, M.: Two patterns of correlated X-ray timing and spectral behaviour in low-mass X-ray binaries,  A\&A, \textbf{225}, 79 (1989)

\bibitem{} Ho L.: On the Relationship between Radio Emission and Black Hole Mass in Galactic Nuclei. ApJ, \textbf{564}, 120 (2002)

\bibitem{} Hynes R.I., Haswell C.A., Cui W., Shrader C.R., O\' Brien K., Chaty S., Skillman D.R., Patterson J., Horne K.: The remarkable rapid X-ray, ultraviolet, optical and infrared variability in the black hole XTE J1118+480. MNRAS, \textbf{345}, 292 (2003)

\bibitem{} Ichimaru S.: Bimodal behavior of accretion disks - Theory and application to Cygnus X-1 transitions. ApJ, 214, 840 (1977)

\bibitem{}Kallman T., Bautista M.:   Photoionization and High-Density Gas,  ApJS,  \textbf{133},  221,  (2001) 

\bibitem{} K\"ording E.G., Fender R.P., Migliari S.: Jet-dominated advective systems: radio and X-ray luminosity dependence on the accretion rate. MNRAS, \textbf{369}, 1451 (2006a)

\bibitem{} K\"ording E.G., Jester S., Fender R.: Accretion states and radio loudness in active galactic nuclei: analogies with X-ray binaries. MNRAS, \textbf{372}, 1366 (2006b)

\bibitem{}K\"ording E., Rupen M., Knigge C., Fender R., Dhawan V., Templeton M., Muxlow T.:   A Transient Radio Jet in an Erupting Dwarf Nova,  Sci,  \textbf{320},  1318,  (2008)

\bibitem{2007ApJ...667.1073L} Lin, D., Remillard, R.~A., 
\& Homan, J.: Evaluating Spectral Models and the X-Ray States of Neutron Star X-Ray Transients. ApJ, \textbf{667}, 1073 (2007)

\bibitem{2009ApJ...696.1257L} Lin, D., Remillard, R.~A., 
\& Homan, J.: Spectral States of XTE J1701 - 462: Link Between Z and Atoll Sources. ApJ, \textbf{696}, 1257 (2009)

\bibitem{}Lee J. C., Reynolds C. S., Remillard R., Schulz N. S., Blackman E. G., Fabian A. C.:   High-Resolution Chandra HETGS and Rossi X-Ray Timing Explorer Observations of GRS 1915+105: A Hot Disk Atmosphere and Cold Gas Enriched in Iron and Silicon,  ApJ,  \textbf{567},  1102,  (2002) 

\bibitem{} Lin D.N.C., Papaloizou J.C.B.: Theory of Accretion Disks II: Application to Observed Systems. ARA\&A, 34, 703 (1996)

\bibitem{} Lohfink A.M. et a.: An X-Ray View of the Jet Cycle in the Radio-loud AGN 3C120. ApJ, \textbf{772}, L83 (2013)

\bibitem{}Luketic S., Proga D., Kallman T. R., Raymond J. C., Miller J. M.:   On the Properties of Thermal Disk Winds in X-ray Transient Sources: A Case Study of GRO J1655-40,  ApJ,  \textbf{719},  515,  (2010)

\bibitem{} Mahadevan R.: Scaling Laws for Advection-dominated Flows: Applications to Low-Luminosity Galactic Nuclei. ApJ, \textbf{477}, 585 (1997)

\bibitem{} McHardy I., K\"ording E., Knigge C., Uttley P., Fender R.: Active galactic nuclei as scaled-up Galactic black holes. Nature, \textbf{444}, 730 (2006)

\bibitem{}McClintock J. E., Narayan R., Steiner J. F.:   Black Hole Spin via Continuum Fitting and the Role of Spin in Powering Transient Jets,  SSRv,  \textbf{183},  295,  (2014)

\bibitem{} McNamara B.R., Nulsen P.E.J.:Heating Hot Atmospheres with Active Galactic Nuclei. ARA\&A, \textbf{45}, 117 (2007)

\bibitem{} Marscher A.P., Jorstad S., Gomez J.L., Aller M.F., Terasranta H., Lister M.L., Stirling A.M.:Observational evidence for the accretion-disk origin for a radio jet in an active galaxy. Nature, \textbf{417}, 625 (2002) 

\bibitem{}Migliari S., Fender R. P.:   Jets in neutron star X-ray binaries: a comparison with black holes,  MNRAS,  \textbf{366},  79,  (2006)

\bibitem{}Migliari S., Fender R. P., Rupen M., Jonker P. G., Klein-Wolt M., Hjellming R. M., van der Klis M.:   Disc-jet coupling in an atoll-type neutron star X-ray binary: 4U 1728-34 (GX 354-0),  MNRAS,  \textbf{342},  L67,  (2003)

\bibitem{} Miller J.M.: ADD TITLE
ARA\&A, 45, 441

\bibitem{} Miller-Jones J.C.A., Moin A., Tingay S.J., Reynolds C., Phillips C.J., Tziounmis A.K., Fender R.P., McCallum J.N., Nicolson G.D., Tudose V.: The first resolved imaging of milliarcsecond-scale jets in Circinus X-1. MNRAS, \textbf{419}, 49 (2012a)

\bibitem{} Miller-Jones J.C.A. {\em et al.}: Disc-jet coupling in the 2009 outburst of the black hole candidate H1743-322. MNRAS, \textbf{421}, 468 (2012b)

\bibitem{} Mirabel I.F., Rodriguez L.F.: A superluminal source in the Galaxy. Nature, \textbf{371}, 46 (1994)

\bibitem{1989PASJ...41...97M} Mitsuda, K., Inoue, H., 
Nakamura, N., \& Tanaka, Y.: Luminosity-related changes of the energy spectrum of X1608-522, PASJ,\textbf{41}, 97 (1989)

\bibitem{}Motta S. E., Belloni T. M., Stella L., Mu\~noz-Darias T., Fender R.:   Precise mass and spin measurements for a stellar-mass black hole through X-ray timing: the case of GRO J1655-40,  MNRAS,  \textbf{437},  2554,  (2014a) 

\bibitem{}Motta S. E., Mu\~noz-Darias T., Sanna A., Fender R., Belloni T., Stella L.:   Black hole spin measurements through the relativistic precession model: XTE J1550-564,  MNRAS,  \textbf{439},  L65,  (2014b)

\bibitem{2014MNRAS.443.3270M} Mu{\~n}oz-Darias, T., Fender, R.~P., Motta, S.~E., 
\& Belloni, T.~M.: Black hole-like hysteresis and accretion states in neutron star low-mass X-ray binaries. MNRAS, \textbf{443}, 3270 (2014)

\bibitem{2011MNRAS.410..679M} 
Mu{\~n}oz-Darias, T., Motta, S., \& Belloni, T.~M.: Fast variability as a tracer of accretion regimes in black hole transients. MNRAS, \textbf{410}, 679 (2011)

\bibitem{}Miller J. M., Raymond J., Fabian A., Steeghs D., Homan J., Reynolds C., van der Klis M., Wijnands R.:   The magnetic nature of disk accretion onto black holes,  Natur,  \textbf{441},  953,  (2006)

\bibitem{} Neilsen J., Lee J.C., 2009. Accretion disk winds as the jet suppression mechanism in the microquasar GRS 1915+105. Nature, \textbf{458}, 481 (2009)

\bibitem{}Neilsen J., Remillard R. A., Lee J. C.:   The Physics of the "Heartbeat" State of GRS 1915+105,  ApJ,  \textbf{737},  69,  (2011) 

\bibitem{}\"{O}zel D., Psaltis D., Narayan R., McClintock J.: The Black Hole Mass Distribution in the Galaxy. ApJ, 725, 1918 (2010)

\bibitem{} Papaloizou J.C.B., Lin D.N.C.: Theory Of Accretion Disks I: Angular Momentum Transport Processes. ARA\&A 33, 505 (1995)

\bibitem{}Plant D. S., Fender R. P., Ponti G., Mu\~noz-Darias T., Coriat M.:   Revealing accretion on to black holes: X-ray reflection throughout three outbursts of GX 339-4,  MNRAS,  \textbf{442},  1767,  (2014) 

\bibitem{} Plotkin R., Markoff S., Kelly B.C., K`"ording E., Anderson S.F.: Using the Fundamental Plane of black hole activity to distinguish X-ray processes from weakly accreting black holes. MNRAS, \textbf{419}, 276 (2012)

\bibitem{}Ponti G., Bianchi S., Mu\~noz-Darias T., De Marco B., Dwelly T., Fender R. P., Nandra K., Rea N., Mori K., Haggard D., Heinke C. O., Degenaar N., Aramaki T., Clavel M., Goldwurm A., Hailey C. J., Israel G. L., Morris M. R., Rushton A., Terrier R.:   On the Fe K absorption - accretion state connection in the Galactic Centre neutron star X-ray binary AX J1745.6-2901,  MNRAS,  \textbf{446},  1536,  (2015)

\bibitem{}Ponti G., Fender R. P., Begelman M. C., Dunn R. J. H., Neilsen J., Coriat M.:   Ubiquitous equatorial accretion disc winds in black hole soft states,  MNRAS,  \textbf{422},  L11,  (2012) 

\bibitem{}Ponti G., Mu\~noz-Darias T., Fender R. P.:   A connection between accretion state and Fe K absorption in an accreting neutron star: black hole-like soft-state winds?,  MNRAS,  \textbf{444},  1829,  (2014) 

\bibitem{} Pringle J.E., Rees M.J.: Accretion Disc Models for Compact X-Ray Sources. A\&A 21, 1 (1972)

\bibitem{}Proga D., Kallman T. R.:   On the Role of the Ultraviolet and X-Ray Radiation in Driving a Disk Wind in X-Ray Binaries,  ApJ,  \textbf{565},  455,  (2002) 

\bibitem{} Psaltis D.:Probes and Tests of Strong-Field Gravity with Observations in the Electromagnetic Spectrum. Living Reviews in Relativity, \textbf{11}, 9 (2008)

\bibitem{} Rees M.J., Begelman M.C., Blandford R.D., Phinney E.S.: Ion-supported tori and the origin of radio jets. Nature, 295, 17 (1982)

\bibitem{}Reynolds C. S.:   Measuring Black Hole Spin Using X-Ray Reflection Spectroscopy,  SSRv,  \textbf{183},  277  (2014)

\bibitem{} Russell D.M., Maccarone T.J., K\"ording E.G., Homan J.: Parallel tracks in infrared versus X-ray emission in black hole X-ray transient outbursts: a hysteresis effect?. MNRAS, \textbf{379}, 1401 (2007)

\bibitem{} Russell D.M., Miller-Jones J.C.A., Maccarone T.J., Yang Y.J., Fender R.P., Lewis F.: Testing the Jet Quenching Paradigm with an Ultradeep Observation of a Steadily Soft State Black Hole. MNRAS, \textbf{739}, L19, 2011

\bibitem{} Steiner J.F., McClintock J.E., Narayan R.: ADD TITLE
ApJ, {\bf 762}, 104 (2013)

\bibitem{} Shakura N.I., Sunyaev R.A.: Black holes in binary systems. Observational appearance. A\&A, \textbf{24}, 337 (1973)

\bibitem{}
Soltan A.: Masses of quasars. MNRAS, \textbf{200}, 115 (1982)

\bibitem{1986SvAL...12..117S} Sunyaev, R.~A., \& Shakura, N.~I.: Disk Accretion onto a Weak Field Neutron Star - Boundary Layer Disk Luminosity Ratio . Soviet Astronomy Letters, \textbf{12}, 117 (1986)

\bibitem{}Tarter C. B., Tucker W. H., Salpeter E. E.:   The Interaction of X-Ray Sources with Optically Thin Environments,  ApJ,  \textbf{156},  943,  (1969) 

\bibitem{2006A&A...454..559Y} Yungelson, L.~R., Lasota, J.-P., Nelemans, G., et al.: The origin and fate of short-period low-mass black-hole binaries. A\&A, \textbf{454}, 559 (2006) 

\bibitem{2006csxs.book...39V} van der Klis, M.: Rapid X-ray Variability.  
Compact stellar X-ray sources, Cambridge Astrophysics Series, No. \textbf{39} (2006)

\bibitem{1988ApJ...324..363W} White, N.~E., Stella, L., 
\& Parmar, A.~N.: The X-ray spectral properties of accretion discs in X-ray binaries. ApJ, \textbf{324}, 363 (1988)

\bibitem{}Wijnands R., van der Klis M.:   The Broadband Power Spectra of X-Ray Binaries,  ApJ,  \textbf{514},  939,  (1999)

\bibitem{}Zdziarski A.A.: The jet kinetic power, distance and inclination of GRS 1915+105. MNRAS, \textbf{444}, 1113 (2014a)

\bibitem{}Zdziarski A.A.: The minimum jet power and equipartition. MNRAS. \textbf{445}, 1321 (2014b)

\bibitem{} Zhang S.N., Cui W., Harmon B.A., Paciesas W.S., Remillard R.E., van Paradijs J.: 
The 1996 Soft State Transition of Cygnus X-1. ApJ, \textbf{477}, L95 (1997)
\end{thebibliography}
\end{document}